\documentclass[12pt,epsf]{article}

\usepackage{times}
\usepackage{graphicx}
\usepackage{amsfonts}
\usepackage{amsmath}
\usepackage{amssymb}
\usepackage{amstext}
\usepackage{amsthm}
\usepackage{mathtools}

\usepackage{epsfig}
\usepackage{color}
\psfigdriver{dvips}
\usepackage{latexsym}
\usepackage[matrix,curve]{xypic}
\usepackage{rotating}
\rotdriver{dvips}
\usepackage{diagrams}

\begin{document}

\baselineskip 6mm
\renewcommand{\thefootnote}{\fnsymbol{footnote}}


\newcommand{\nc}{\newcommand}
\newcommand{\rnc}{\renewcommand}


\rnc{\baselinestretch}{1.24}    
\setlength{\jot}{6pt}       
\rnc{\arraystretch}{1.24}   

\makeatletter
\rnc{\theequation}{\thesection.\arabic{equation}}
\@addtoreset{equation}{section}
\makeatother



\nc{\be}{\begin{equation}}

\nc{\ee}{\end{equation}}

\nc{\bea}{\begin{eqnarray}}

\nc{\eea}{\end{eqnarray}}

\nc{\xx}{\nonumber\\}

\nc{\ct}{\cite}

\nc{\la}{\label}

\nc{\eq}[1]{(\ref{#1})}

\nc{\newcaption}[1]{\centerline{\parbox{6in}{\caption{#1}}}}

\nc{\fig}[3]{

\begin{figure}
\centerline{\epsfxsize=#1\epsfbox{#2.eps}}
\newcaption{#3. \label{#2}}
\end{figure}
}


\def\CA{{\cal A}}
\def\CC{{\cal C}}
\def\CD{{\cal D}}
\def\CE{{\cal E}}
\def\CF{{\cal F}}
\def\CG{{\cal G}}
\def\CH{{\cal H}}
\def\CK{{\cal K}}
\def\CL{{\cal L}}
\def\CM{{\cal M}}
\def\CN{{\cal N}}
\def\CO{{\cal O}}
\def\CP{{\cal P}}
\def\CR{{\cal R}}
\def\CS{{\cal S}}
\def\CU{{\cal U}}
\def\CV{{\cal V}}
\def\CW{{\cal W}}
\def\CY{{\cal Y}}
\def\CZ{{\cal Z}}


\def\IB{{\hbox{{\rm I}\kern-.2em\hbox{\rm B}}}}
\def\IC{\,\,{\hbox{{\rm I}\kern-.50em\hbox{\bf C}}}}
\def\ID{{\hbox{{\rm I}\kern-.2em\hbox{\rm D}}}}
\def\IF{{\hbox{{\rm I}\kern-.2em\hbox{\rm F}}}}
\def\IH{{\hbox{{\rm I}\kern-.2em\hbox{\rm H}}}}
\def\IN{{\hbox{{\rm I}\kern-.2em\hbox{\rm N}}}}
\def\IP{{\hbox{{\rm I}\kern-.2em\hbox{\rm P}}}}
\def\IR{{\hbox{{\rm I}\kern-.2em\hbox{\rm R}}}}
\def\IZ{{\hbox{{\rm Z}\kern-.4em\hbox{\rm Z}}}}


\def\a{\alpha}
\def\b{\beta}
\def\d{\delta}
\def\ep{\epsilon}
\def\ga{\gamma}
\def\k{\kappa}
\def\l{\lambda}
\def\s{\sigma}
\def\t{\theta}
\def\w{\omega}
\def\G{\Gamma}


\def\half{\frac{1}{2}}
\def\dint#1#2{\int\limits_{#1}^{#2}}
\def\goto{\rightarrow}
\def\para{\parallel}
\def\brac#1{\langle #1 \rangle}
\def\curl{\nabla\times}
\def\div{\nabla\cdot}
\def\p{\partial}


\def\Tr{{\rm Tr}\,}
\def\det{{\rm det}}


\def\vare{\varepsilon}
\def\zbar{\bar{z}}
\def\wbar{\bar{w}}
\def\what#1{\widehat{#1}}


\def\ad{\dot{a}}
\def\bd{\dot{b}}
\def\cd{\dot{c}}
\def\dd{\dot{d}}
\def\so{SO(4)}
\def\bfr{{\bf R}}
\def\bfc{{\bf C}}
\def\bfz{{\bf Z}}

\begin{titlepage}


\hfill\parbox{3.7cm} {{\tt arXiv:1804.09171}}

\vspace{15mm}

\begin{center}
{\Large \bf  Quantized K\"ahler Geometry and Quantum Gravity}

\vspace{10mm}

Jungjai Lee ${}^{a}$\footnote{jjlee@daejin.ac.kr} and Hyun Seok Yang ${}^b$\footnote{hsyang@sogang.ac.kr}
\\[10mm]

${}^a$ {\sl Division of Mathematics and Physics, Daejin University, Pocheon, Gyeonggi 487-711, Korea}

${}^b$ {\sl Center for Quantum Spacetime, Sogang University, Seoul 04107, Korea}

\end{center}

\thispagestyle{empty}

\vskip1cm


\centerline{\bf ABSTRACT}
\vskip 4mm
\noindent

It has been often observed that K\"ahler geometry is essentially a $U(1)$ gauge theory
whose field strength is identified with the K\"ahler form.
However it has been pursued neither seriously nor deeply.
We argue that this remarkable connection between the K\"ahler geometry and $U(1)$ gauge theory
is a missing corner in our understanding of quantum gravity.
We show that the K\"ahler geometry can be described by a $U(1)$ gauge theory on a symplectic manifold
with a slight generalization. We derive a natural Poisson algebra associated with the K\"ahler geometry
we have started with. The quantization of the underlying Poisson algebra leads to a noncommutative
$U(1)$ gauge theory which arguably describes a quantized K\"ahler geometry.
The Hilbert space representation of quantized K\"ahler geometry eventually ends in a zero-dimensional matrix model.
We then play with the zero-dimensional matrix model to examine how to recover
our starting point--K\"ahler geometry--from the background-independent formulation.
The round-trip journey suggests many remarkable pictures for quantum gravity
that will open a new perspective to resolve the notorious problems in theoretical physics such as
the cosmological constant problem, hierarchy problem, dark energy, dark matter and cosmic
inflation. We also discuss how time emerges to generate a Lorentzian spacetime
in the context of emergent gravity.
\\


Keywords: Quantization of K\"ahler geometry, Noncommutative $U(1)$ gauge theory, Matrix model,
Quantum gravity, Emergent gravity

\vspace{1cm}

\today

\end{titlepage}

\renewcommand{\thefootnote}{\arabic{footnote}}
\setcounter{footnote}{0}

\section{Introduction}

The concept of emergent gravity and spacetime recently activated by the AdS/CFT correspondence
advocates that spacetime is not a fundamental entity existed from the beginning but an emergent
property from something much deeper \cite{est-review,witten-se}.
However, the ``emergence" here means the emergence not just of the gravitational field
but of the spacetime on which the gravitational field propagates.
Any emergent theory of gravity should have this property, since an essential part of gravity is
that spacetime is free to fluctuate and cannot be built in from the beginning.
Thus, in order to realize the concept of emergent spacetime, the spacetime must be replaced by
a primal monad such as matrices and it has to be derived in some limit from the deeper structure.
If the spacetime we experience is emergent from something deeper, the particles and fields in it
should be all emergent too from the same structure because they are some structures supported
on the spacetime. This implies that emergent spacetime also enforces emergent quantum mechanics.
In other words, emergent spacetime requires to unify geometry and matter so that
spacetime and matter are emergent together from a universal structure in microscopic level.

The important question is then what is the universal structure in microscopic level to realize
emergent spacetime as well as emergent quantum mechanics. Of course, the universal structure
must be the crux for quantum gravity. Recall that quantum mechanics has already shown us
such a radical change in physics \cite{dirac}.
Quantum mechanics is the formulation of mechanics on noncommutative (NC)
phase space whose coordinate generators satisfy the commutation relation given by
\begin{equation}\label{nc-phase}
[ x^i, p^j ]= i \hbar\delta^{ij}.
\end{equation}
Since its advent, quantum mechanics constantly teaches us that $\hbar$ is not a word.
We are still debating on the quantum reality. In this sense, we may be ready to accept
such a revolutionary change from quantum gravity. Hence the novel idea such as emergent spacetime
just made must not be hindered by too restricted concepts, and the progress in comprehending
the connection of things should not be obstructed by traditional prejudices.

Quantum mechanics defines a more fundamental principle and general framework than classical mechanics,
so the latter is derived (or emergent) from the former in a limit $\hbar \to 0$.
Similarly, quantum gravity is expected to define a more fundamental principle and general framework
than general relativity, so the latter may be derived (or emergent) from the former in a limit $G \to 0$.
Here $G$ is the gravitational constant which specifies a certain scale $l_P = \sqrt{\frac{G\hbar}{c^3}}
= 10^{-33}$ cm where quantum gravity becomes important.
A natural reasoning is thus that the universal structure for quantum gravity would be
related to a new physics appearing when $G$ plays a crucial role.
The dimensional argument implies that spacetime at the Planck length $l_P$ is no longer commuting,
instead spacetime coordinates obey the commutation relation \cite{snyder-yang}
\begin{equation}\label{nc-space-int}
    [y^\mu, y^\nu] = i \theta^{\mu\nu}.
\end{equation}
The Heisenberg algebra generated by the coordinate generators $y^\mu \; (\mu = 1, \cdots, 2n)$
will be denoted by $\mathbb{R}_\theta^{2n}$ and the NC algebra defined on $\mathbb{R}_\theta^{2n}$
by $\mathcal{A}_\theta$.
To be specific, let us take the $2n \times 2n$ symplectic matrix $\theta^{\mu\nu}$ as the form
\begin{equation} \label{symp-theta}
    \theta^{\mu\nu} = \alpha' \left(
                            \begin{matrix}
\varepsilon &  0  & \ldots & 0 \\
0  &  \varepsilon & \ldots & 0 \\
\vdots & \vdots & \ddots & \vdots\\
0  &   0       &\ldots & \varepsilon
\end{matrix}
                          \right)
\end{equation}
where $\alpha'\equiv l_s^2$ is a fundamental constant with the physical dimension of $(\mathrm{length})^2$
and $\varepsilon = i \sigma^2 $ is the $2 \times 2$ symplectic matrix.
Given a polarization like \eq{symp-theta}, it is convenient to split the coordinate generators
as $y^\mu = (y^{2i-1}, y^{2i}), \;  i=1,\cdots,n$, and rename them as $y^{2i-1} \equiv x^i$ and
$y^{2i} \equiv \frac{\alpha'}{\hbar} p^i$.
We have intentionally introduced the Planck constant $\hbar$.
Note that $[y^{2i}]$ carries the physical dimension of length, as it should be, if $p^i$ is a momentum.
Then the commutation relation \eq{nc-space-int} can be written as the form \eq{nc-phase}.
Therefore we can apply the Heisenberg uncertainty principle to the commutation relation \eq{nc-phase}
which leads to
\begin{equation}\label{uncertain-qm}
    \Delta x^i  \Delta p^j  \geq \frac{\hbar}{2} \delta^{ij}.
\end{equation}
If we use the original variables $y^\mu = (y^{2i-1}, y^{2i})$, the above uncertainty relation reads as
\begin{equation}\label{uncertain-nc}
    \Delta y^{2i-1} \Delta y^{2j}  \geq \frac{\alpha'}{2} \delta^{ij}.
\end{equation}

A trivial but allusive fact is that the mathematical structure of NC space is the same
as quantum mechanics. Thus one may regard the physics on the NC space \eq{nc-space-int} as a `quantum mechanics'
defined by $\alpha'$ instead of $\hbar$ \cite{hsy-review}. This is the reason why one should not consider
the NC space $\mathbb{R}_\theta^{2n}$ as a classical space. Indeed we can learn every
important lessons from quantum mechanics \cite{dirac}:

$\mathfrak{A}$. NC space $\mathbb{R}_\theta^{2n}$ introduces a separable Hilbert space $\mathcal{H}$ and
dynamical variables in $\mathcal{A}_\theta$ become operators acting on the Hilbert space $\mathcal{H}$.

$\mathfrak{B}$. NC algebra $\mathcal{A}_\theta$ admits a nontrivial inner automorphism given by
\begin{equation}\label{inn-auto}
    \widehat{f}(y+d) = U \widehat{f}(y) U^{-1}
\end{equation}
for $\widehat{f}(y) \in \mathcal{A}_\theta$ and $U = \exp (i y^\mu \theta^{-1}_{\mu\nu} d^\nu)$.
This implies that every points in the NC space $\mathbb{R}_\theta^{2n}$ are unitarily equivalent.
Thus the concept of classical space(time) is doomed and the space(time) is replaced by
a quantum algebra $(\mathcal{H}, \mathcal{A}_\theta)$. A classical spacetime is derived (or emergent)
from the quantum algebra in a specific limit.

$\mathfrak{C}$. A dynamical field becomes a linear operator acting on the Hilbert space $\mathcal{H}$ and
any linear operator is represented by a matrix. The matrix representing the product of two linear operators
is the product of the matrices representing the two factors. This means that a theory of dynamical fields on $\mathbb{R}_\theta^{2n}$ eventually reduces to a matrix model.

The lessons $(\mathfrak{A}, \mathfrak{B}, \mathfrak{C})$ are enough to draw the conclusion that
NC spacetime necessarily implies
emergent spacetime if spacetime at microscopic scales should be viewed as NC \cite{q-emg,hsy-2016}.
It is striking to see how quantum mechanics provides us the underlying idea that
the spacetime we live in and all the particles and forces in it must be emergent in a consistent way
with the modern understanding of quantum gravity.
It turns out \cite{hsy-2016,dmde} that the emergent spacetime is a new fundamental paradigm that
allows a background-independent formulation of quantum gravity and opens a new perspective
to resolve the notorious problems in theoretical physics
such as the cosmological constant problem, hierarchy problem, dark energy, and dark matter.
Furthermore, the emergent spacetime picture admits a background-independent description of
the inflationary universe which has a sufficiently elegant and explanatory power to defend the integrity
of physics against the multiverse hypothesis \cite{hsy-inflation,hsy-essay}.

Moreover emergent spacetime seems to be much more radical than quantum mechanics for the following reason.
The scales where the NC (or quantum) effect becomes significant
are dramatically different for the NC space and quantum mechanics. Since the noncommutativity of spacetime is set
by the fundamental constant $\alpha' = l_s^2$, it is natural to consider the length scale as
the Planck length, i.e. $l_s= 10^{-35}$m. $l_s$ is much more smaller than the scale for quantum mechanics,
typically the Bohr radius $r_B = 5.3 \times 10^{-11}$m.
The Bohr radius corresponds to the size of superclusters ($\sim 10^{24}$m) in our Universe to an observer
who appreciates the NC effect \eq{uncertain-nc} occurring in the NC space $\mathbb{R}_\theta^{2n}$.
So far, the smallest distance accessible in experiments on Earth is about 10$^{-20}$m at the LHC.
Even this scale amounts to the remote boundary, the Oort Cloud, of the Solar system to the Planckian observer.
Hence the quantum mechanics represented by the NC phase space \eq{nc-phase} rather behaves like
a ``classical system" to an observer near the Planck scale.
Hence it may be reasonable to consider quantum mechanics as equally emergent from a primal monad underlying
the emergent spacetime. The NC space \eq{nc-space-int} is arguably such a primal monad
since it is a primitive vacuum algebra responsible for the generation of space and time.

The space uncertainty relation \eq{uncertain-nc} implies the UV/IR mixing that small scale (UV) fluctuations
are paired with large scale (IR) fluctuations.
Although the UV/IR mixing was derived in \cite{uv-ir} from quantum loops controlled by $\hbar$,
it is obvious from the uncertainty relation \eq{uncertain-nc} that the UV/IR mixing should exist even without
considering quantum mechanics, i.e. $\hbar$-effects, since the NC space \eq{nc-space-int} can be written
as the form \eq{nc-phase}.
Therefore it is necessary to take the uncertainty relation \eq{uncertain-nc} into account as a primary effect
when we consider the physics on the NC space \eq{nc-space-int}.

This paper is organized as follows.
In section 2, we explain how K\"ahler geometry can be described by a $U(1)$ gauge theory
on a symplectic manifold and derive a natural Poisson algebra associated with
the K\"ahler geometry we have started with.
In section 3, we quantize the underlying Poisson algebra to get a NC $U(1)$ gauge theory
which arguably describes a quantized K\"ahler geometry.
The Hilbert space representation of quantized K\"ahler geometry results in a zero-dimensional matrix model.
Hence we support the conjecture in \cite{inov} that
NC $U(1)$ gauge theory is the fundamental description of K\"ahler gravity at all scales including
the Planck scale and provides a quantum gravity description such as quantum gravitational foams.
The duality in \cite{inov} has been further clarified in \cite{neova-kap} by showing that it follows
from the S-duality of the type IIB superstring.
In section 4, we play with the zero-dimensional matrix model to examine how to recover
our starting point--K\"ahler geometry--from the background-independent formulation.
The round-trip journey suggests many remarkable pictures for quantum gravity
that would be significant to resolve the notorious problems in theoretical physics.
Section 5 is devoted to several generalizations going beyond K\"ahler manifolds.
In particular, we discuss how time emerges to generate Lorentzian manifolds as a dynamical spacetime
in general relativity and argue that the quantization of Lorentzian manifolds is described by
the BFSS-type matrix model \cite{q-emg,hsy-inflation,hsy-jhep09}.

We hope that this review contributes to uncovering the dormant picture on
the remarkable connection between the K\"ahler geometry and $U(1)$ gauge theory,
and sheds light on a missing corner in our understanding of quantum gravity.

\section{K\"ahler Geometry and $U(1)$ Gauge Theory}

Let $M$ be an $n$-dimensional complex manifold with a Hermitian metric.
This means that the metric $g$ has only $(1,1)$-type for a given complex structure,
in which $(2,0)$- and $(0,2)$-types are projected out.
In terms of local complex coordinates, $z^i = y^{2i-1} + \sqrt{-1} y^{2i},
\; \overline{z}^{\bar{i}} = y^{2i-1} - \sqrt{-1} y^{2i}, \; (i, \bar{i} = 1, \cdots, n)$,
the metric on $M$ is given by
\begin{equation}\label{c-metric}
    ds^2 = g_{i\bar{j}} (z, \overline{z})  dz^i d \overline{z}^{\bar{j}}.
\end{equation}
Given the Hermitian metric \eq{c-metric}, one can introduce a fundamental two-form defined by
\begin{equation}\label{f2-form}
    \Omega = \sqrt{-1} g_{i\bar{j}} (z, \overline{z})  dz^i \wedge d \overline{z}^{\bar{j}}.
\end{equation}
A K\"ahler manifold is then defined as a Hermitian manifold with the closed fundamental two-form,
i.e., $d\Omega = 0$ \cite{besse}.
The K\"ahler condition is equivalent to the local existence of some function
$K(z, \overline{z})$ such that
\begin{equation}\label{k-metric}
 g_{i\bar{j}} (z, \overline{z}) = \frac{\partial^2  K(z, \overline{z})}{\partial z^i \partial \overline{z}^{\bar{j}}}.
\end{equation}
The function $K(z, \overline{z})$ is called K\"ahler potential. The K\"ahler potential is not unique
but admits a K\"ahler transformation given by
\begin{equation}\label{kah-gauge}
K(z, \overline{z}) \to K(z, \overline{z}) + f(z) + \overline{f}(\overline{z})
\end{equation}
where $f(z)$ and $\overline{f}(\overline{z})$ are arbitrary holomorphic and anti-holomorphic functions.
Two K\"ahler potentials related by the K\"ahler gauge transformation \eq{kah-gauge} give rise
to the same K\"ahler metric \eq{k-metric}. Note that the K\"ahler form \eq{f2-form} can be written as
\begin{equation}\label{k-curvature}
    \Omega = d \mathcal{A} \qquad \mathrm{and} \qquad
    \mathcal{A} = \frac{\sqrt{-1}}{2} (\overline{\partial} - \partial) K (z, \overline{z})
\end{equation}
where the exterior differential operator is given by $d = \partial + \overline{\partial}$ with
$\partial = dz^i \frac{\partial}{\partial z^i}$ and $ \overline{\partial} = d \overline{z}^{\bar{i}} \frac{\partial}{\partial \overline{z}^{\bar{i}}}$.
Then the above K\"ahler transformation \eq{kah-gauge} corresponds to a gauge transformation
for the one-form $\mathcal{A}$ given by
\begin{equation}\label{k-gauge}
    \mathcal{A} \to \mathcal{A} + d \lambda
\end{equation}
where $\lambda = \frac{\sqrt{-1}}{2} \big(\overline{f}(\overline{z}) - f(z) \big)$.
This implies that the one-form $\mathcal{A}$ corresponds to $U(1)$ gauge fields.

The K\"ahler form $\Omega$ on a K\"ahler manifold $M$ is a nondegenerate, closed two-form.
Therefore the K\"ahler form $\Omega$ is a symplectic two-form.
This means that a K\"ahler manifold $(M, \Omega)$ is a symplectic manifold too
although the reverse is not necessarily true. Let us consider an atlas
$\{(U_\alpha, \varphi_\alpha)| \alpha \in I \}$ on
the K\"ahler manifold $M$ and denote the K\"ahler form $\Omega$ restricted
on a chart $(U_\alpha, \varphi_\alpha)$ as $\mathcal{F}_\alpha \equiv \Omega|_{U_\alpha}$.
It is possible to write the local K\"ahler form as\footnote{The K\"ahler condition enforces
a specific analytic characterization of K\"ahler metrics \cite{griffiths-harris}.
For a Hermitian metric $g$ on a complex manifold $(M, J)$, $g$ is K\"ahler
if and only if around each point of $M$, there exist holomorphic coordinates in which $g$
osculates to order 2 to the Euclidean metric on $\mathbb{C}^n$.
This means that the existence of normal holomorphic coordinates around each
point of $M$ is equivalent to that of K\"ahler metrics.}
\begin{equation}\label{kahler-gauge}
\mathcal{F}_\alpha = B + F_\alpha,
\end{equation}
where $B$ is the K\"ahler form of $\mathbb{C}^n$.
Since the two-form $F_\alpha$ must be closed due to the K\"ahler condition,
it can be represented by $F_\alpha = d A_\alpha$. 
Using Eq. \eq{k-curvature} and $F_\alpha = \mathcal{F}_\alpha - B$,
the one-form $A_\alpha$ on $U_\alpha$ can be written as the form
\begin{equation}\label{local-one-form}
    A_\alpha = \frac{\sqrt{-1}}{2} (\overline{\partial} - \partial) \phi_\alpha (z, \overline{z})
\end{equation}
where $\phi_\alpha (z, \overline{z}) = K_\alpha (z, \overline{z}) - K_0 (z, \overline{z})$ and
$K_\alpha (z, \overline{z})$ is the K\"ahler potential on a local chart $U_\alpha$ and $K_0 (z, \overline{z})
= z^i \overline{z}^{\bar{i}}$ is the K\"ahler potential of $\mathbb{C}^n$.
On an overlap $U_\alpha \bigcap U_\beta$, two one-forms $A_\alpha$ and $A_\beta$ can be glued
using the freedom \eq{k-gauge} such that
\begin{equation}\label{g-transf}
    A_\beta = A_\alpha + d \lambda_{\alpha\beta}
\end{equation}
where $\lambda_{\alpha\beta}(z, \overline{z})$ is a smooth function on the overlap $U_\alpha \bigcap U_\beta$.
The gluing \eq{g-transf} on $U_\alpha \bigcap U_\beta$ is equal to the K\"ahler transformation
\begin{equation}\label{holo-gauge}
K_\beta (z, \overline{z}) = K_\alpha (z, \overline{z})
+ f_{\alpha\beta} (z) + \overline{f}_{\alpha\beta} (\overline{z})
\end{equation}
if $\lambda_{\alpha\beta}(z, \overline{z}) = \frac{\sqrt{-1}}{2} \big(\overline{f}_{\alpha\beta}(\overline{z})
- f_{\alpha\beta}(z) \big)$.

These aspects of K\"ahler geometry we have described so far imply that K\"ahler gravity can be
described by a $U(1)$ gauge theory in which the one-form \eq{local-one-form} plays a role of
the connection of a holomorphic line bundle $L$. We will show that the connection between
the K\"ahler gravity and a $U(1)$ gauge theory is remarkably true with a slight generalization \cite{hsy-mirror}.
However, this observation is not new. Iqbal et al. have come to a notice in a beautiful paper \cite{inov}
that K\"ahler gravity is essentially described by a $U(1)$ gauge theory. They conclude that,
for topological strings probing K\"ahler manifolds, the $U(1)$ gauge theory is the fundamental description
of gravity at all scales including the Planck scale, where it leads to quantum gravitational foams.

Now we will explain why a K\"ahler geometry can be formulated in terms of a $U(1)$ gauge theory
on a symplectic manifold. In this scheme, the K\"ahler geometry will be derived from the $U(1)$ gauge theory.
Suppose that $(N, B)$ be a $2n$-dimensional symplectic manifold where $B$ is a symplectic two-form, i.e.,
a nondegenerate and closed two-form on $N$. We emphasize that the manifold $N$ differs
from the K\"ahler manifold $M$ even topologically since $N$ would suffer from a topology change
after the resolution of $U(1)$ instanton singularities. For instance, $N = \mathbb{C}^n$
for a non-compact K\"ahler manifold. Let us consider a line bundle $L$ over $N$ whose
connection and curvature are denoted by $A$ and $F=dA$, respectively. Our purpose to introduce
a line bundle $L$ over a symplectic manifold $(N,B)$ is to realize the K\"ahler gravity in terms of
a $U(1)$ gauge theory. For this purpose, the concept of the line bundle $L \to N$ needs to be
generalized in the following way. First, we need to incorporate the structure \eq{kahler-gauge}
into the line bundle $L$. This can be achieved by introducing the so-called $\Lambda$-symmetry
or $B$-field transformation in string theory \cite{pol-book}.
The $B$-field transformation acts on the symplectic
two-form $B$ as well as the connection $A$ of line bundle as follows:
\begin{equation}\label{b-transf}
    (B, A) \to (B - d\Lambda, A + \Lambda),
\end{equation}
where the gauge parameter $\Lambda$ is a one-form in $N$. Note that $(B, A) \to (B, A + d \lambda)$
if $\Lambda = d \lambda$. Thus the ordinary $U(1)$ gauge symmetry is a particular case of
the $\Lambda$-symmetry generated by an exact one-form. Then the gauge invariant under
the $B$-field transformation is $\mathcal{F} = B + F$ which was already appeared in Eq. \eq{kahler-gauge}
as a local K\"ahler form. One who is familiar with string theory may recognize
the $\Lambda$-symmetry \eq{b-transf} since it is realized in the open string action given by
\begin{equation}\label{open-action}
    S = \frac{1}{4 \pi \alpha'} \int_\Sigma |dX|^2 + \int_\Sigma B + \int_{\partial \Sigma} A
\end{equation}
where $X: \Sigma \to N$ is a map from an open string worldsheet $\Sigma$ to an ambient space $N$.
The open string action \eq{open-action} is definitely invariant under the $B$-field transformation \eq{b-transf}.
As a result, low-energy effective theories on $N$ derived from the open string action \eq{open-action}
such as DBI actions depend only on the gauge-invariant combination $\mathcal{F} = B + F$.
The other generalization is that we need to allow singular $U(1)$ gauge fields on $N$
in order to realize the K\"ahler gravity in terms of a $U(1)$ gauge theory since the singularity
of $U(1)$ instantons on the commutative space $N$ is resolved as we will see later.
To admit such a singular gauge field, we need to relax the notion of the line bundle $L$.
The natural replacement for the holomorphic line bundle $L$ is the rank one torsion free sheaf
with the same first Chern class or an ideal sheaf \cite{inov,mnop}.
Note that torsion free sheaves fail to be a line bundle
in real codimension four. We will assume the generalization of the line bundle
by allowing singular $U(1)$ gauge fields at finite number of points and on higher-dimensional cycles.

Note that $\mathcal{F}_\alpha = B + F_\alpha$ in \eq{kahler-gauge} is a K\"ahler form
on a local patch $U_\alpha \subset M$.
Thus it is related to a local metric $g_\alpha$ on $U_\alpha$ by the relation
$g_\alpha (X, Y)= \mathcal{F}_\alpha (X, JY)$ for any vector fields $X, Y \in \Gamma(TM)$
where $J$ is a complex structure on $M$.
Since $B$ is the K\"ahler form of $\mathbb{C}^n$, the local metric can be written as $(g_\alpha)_{i\bar{j}} = \delta_{i\bar{j}} + (h_\alpha)_{i\bar{j}}$ where $h_\alpha (X, Y)= F_\alpha (X, JY)$
describes the local deformations of a background space, $\mathbb{C}^n$ in our case.
The local complex structure $J$ on $U_\alpha \subset M$ is inherited from the
local symplectic structure and it is determined by
\begin{equation}\label{local-comp}
B (X, JY) = \delta (X, Y)     \qquad \Rightarrow \qquad B_{\mu\lambda} {J^\lambda}_\nu =
\delta_{\mu\nu},  \quad \mathrm{for} \; X, Y  \in \Gamma(T U_\alpha),
\end{equation}
where $\delta_{\mu\nu}$ is the flat metric on $\mathbb{C}^n$.
Now we identify the K\"ahler form $\mathcal{F}_\alpha = B + F_\alpha$ in \eq{kahler-gauge} with
a line bundle $L$ over a symplectic manifold $(N, B)$ which respects the $\Lambda$-symmetry \eq{b-transf}.
The curving of a background space is now described by local fluctuations of $U(1)$ gauge fields in
the line bundle $L \to N$. Thus the dynamical $U(1)$ gauge fields defined on a symplectic manifold $(N,B)$
manifest themselves as local deformations of the symplectic or K\"ahler structure and
they correspond to gravitational fields on a background space according to $h_\alpha (X, Y)= F_\alpha (X, JY)$.
In this sense, the underlying symplectic manifold is a dynamical system and locally described by
$(N = \bigcup_\alpha U_\alpha, \mathcal{F}_\alpha = B + F_\alpha)$.\footnote{Here the dynamics simply
refers to local fluctuations around a background configuration.
We will discuss later how the dynamical time evolution of the system can be defined through local deformations.}
An original K\"ahler geometry results in the dynamics of $U(1)$ gauge fields.
The dynamical system is constructed locally on each local chart $(U_\alpha, \varphi_\alpha)$ and
the local construction can be glued using the gauge degrees of freedom \eq{g-transf} or \eq{holo-gauge}.
In this way K\"ahler gravity has a description in terms of $U(1)$ gauge theory.
So the K\"ahler geometry is essentially understood as a dynamical symplectic manifold which carries an
intrinsic Poisson structure $\theta \equiv B^{-1} \in \Gamma(\Lambda^2 TN)$ derived from
the underlying symplectic structure $B$.
We take the Poisson tensor $\theta^{\mu\nu} = \alpha' ( \mathbb{I}_n \otimes i\sigma^2 )$ as
the matrix in \eq{symp-theta}. We have introduced a constant parameter
$\alpha' \equiv l_s^2$ carrying the physical dimension of $\mathrm{(length)}^2$ which will characterize
the fundamental length scale of NC spaces after quantization.
One may try to quantize the dynamical symplectic manifold in the exactly same way as quantum mechanics.
We will show that the quantization of the dynamical symplectic manifold leads to a dynamical
NC space which is described by NC $U(1)$ gauge theory \cite{q-emg}.
In the end we will derive the conclusion from the quantization that the quantized K\"ahler geometry
is described by NC $U(1)$ gauge theory, as conjectured in \cite{inov}.

If K\"ahler gravity can be modeled by a $U(1)$ gauge theory on a symplectic manifold,
it is necessary for the $U(1)$ gauge theory to realize the equivalence principle and diffeomorphism symmetry,
which are arguably the most important properties in the theory of general relativity.
At first sight, it seems to be impossible. Let us explain why the $U(1)$ gauge theory on a symplectic manifold
is radically different from the usual Maxwell's electromagnetism.
First, note that the $\Lambda$-symmetry \eq{b-transf} is possible only when $B \neq 0$.
Eq. \eq{open-action} clearly shows that the $\Lambda$-symmetry is reduced to the ordinary $U(1)$ gauge symmetry,
$A \to A + d \lambda$, if $B = 0$. Therefore the gauge symmetry is rather greatly enhanced whenever
the base space supports a symplectic structure \cite{hsy-ijmp06}.
Moreover, the underlying symplectic structure provides a bundle isomorphism $B: T N \to T^* N$
defined by $X \mapsto A = \iota_X B$ for $X \in \Gamma(TN)$ and $A \in \Gamma (T^* N)$ where
$\iota_X$ is the interior product of vector field $X$.
Since we want to identify the one-form $A$ with the connection of line bundle $L \to N$,
we introduce the equivalence relation given by
\begin{equation}\label{eq-rel}
    X \sim X' = X + X_\lambda
\end{equation}
where $X_\lambda$ is a Hamiltonian vector field defined by $\iota_{X_\lambda} B = d \lambda$.
Then the field strength of $U(1)$ gauge fields is given by $F = dA = d \iota_X B =
(d \iota_X + \iota_X d) B = \mathcal{L}_X B$ where $\mathcal{L}_X$ is the Lie derivative
with respect to vector field $X$. Note that $F = \mathcal{L}_X B = \mathcal{L}_{X'} B$ as it should be.
Consequently the dynamical symplectic two-form $\mathcal{F}$ is locally represented by
\begin{equation}\label{darboux-2form}
    \mathcal{F} = B + F = (1 + \mathcal{L}_X) B \approx e^{\mathcal{L}_X} B.
\end{equation}
Note that vector fields are Lie algebra generators of diffeomorphisms on $N$.
Therefore $e^{\mathcal{L}_X}$ in Eq. \eq{darboux-2form} is equal to some finite coordinate transformation
generated by the vector field $X$ on a local coordinate patch $U_\alpha$ which is denoted
by $\varphi_\alpha \in \mathrm{Diff}(U_\alpha)$. Then Eq. \eq{darboux-2form} implies that
the electromagnetic fields can always be eliminated by a local coordinate transformation.
To be specific, in terms of local coordinates represented by $\varphi_\alpha: y^\mu \mapsto x^a =x^a (y),
\; \mu, a = 1, \cdots, 2n$, Eq. \eq{darboux-2form} means
that $\varphi_\alpha^* (B + F) = B$, i.e., \cite{sw-darboux}
\begin{equation}\label{darboux-tr}
    \Big (B_{ab} + F_{ab} (x) \Big) \frac{\partial x^a}{\partial y^\mu} \frac{\partial x^b}{\partial y^\nu}
    = B_{\mu\nu}.
\end{equation}
Actually this statement is known as the Darboux theorem or Moser lemma
in symplectic geometry \cite{symp-book}.

Let us clarify why the Darboux theorem or Moser lemma in symplectic geometry explains the equivalence principle
in general relativity. As we discussed above, the local K\"ahler form in Eq. \eq{darboux-2form} corresponds
to a local K\"ahler metric $g_{\mu\nu} (x) = \delta_{\mu\nu} + h_{\mu\nu} (x)$, so Eq. \eq{darboux-2form} or \eq{darboux-tr} is equivalently stated as $\varphi_\alpha^* \big(g_{\mu\nu}) = \delta_{\mu\nu}$.
Note that the local complex structure in Eq. \eq{local-comp} can be written as ${J^{\mu}}_\nu= \varepsilon^{\mu\lambda}
\delta_{\lambda\nu}$ where $\varepsilon^{\mu\nu} = \frac{\theta^{\mu\nu}}{\alpha'}$.
Therefore the Darboux theorem or Moser lemma in symplectic geometry corresponds to the equivalence principle
in general relativity. However it should be remarked that there is a crucial difference
between symplectic geometry and general relativity.
The Darboux theorem holds on an entire open neighborhood $U_\alpha \subset N$ \cite{symp-book} whereas
the equivalence principle in general relativity holds only on an infinitesimal neighborhood of a point in $N$.
In other words, the equivalence principle in general relativity may be regarded as the infinitesimal limit
of the Darboux theorem or Moser lemma in symplectic geometry.
In addition, the bundle isomorphism $B: T N \to T^* N$ implies that the $\Lambda$-symmetry \eq{b-transf}
is isomorphic to diffeomorphism symmetry $\mathrm{Diff} (N)$. Indeed this isomorphism is known as
$\beta$-diffeomorphism in generalized geometry \cite{beta-diff}. The important point is that the $B$-field
transformation \eq{b-transf} is also involved with dynamical $U(1)$ gauge fields, so it can be promoted to
dynamical diffeomorphisms, a.k.a. the equivalence principle in general relativity as we have illuminated above.
This is the reason why the K\"ahler geometry can be described by a $U(1)$ gauge theory
as prudentially claimed in \cite{inov}.

Suppose that the coordinate transformation in Eq. \eq{darboux-tr} takes the form
\begin{equation}\label{x-phi}
    \varphi_\alpha: y^\mu \mapsto x^a (y) = y^a + \theta^{ab} a_b(y) =  \theta^{ab} \big( p_b + a_b(y) \big)
    \equiv \theta^{ab} \phi_b (y)
\end{equation}
where $p_a = B_{ab} y^b$.
Let us introduce the Poisson bracket defined by the Poisson tensor
$\theta = B^{-1} \in \Gamma(\Lambda^2 TN)$ as
\begin{equation}\label{p-bracket}
    \{f, g\}_\theta = \theta (df, dg) \qquad \mathrm{or} \qquad
    \{f, g\}_\theta = \theta^{\mu\nu} \frac{\partial f(y)}{\partial y^\mu}\frac{\partial g(y)}{\partial y^\nu}
\end{equation}
for any smooth functions $f, g \in C^\infty (N)$.
Then one can calculate the following Poisson brackets
\begin{eqnarray}\label{poisson-br1}
    && \{y^\mu, y^\nu \}_\theta = \theta^{\mu\nu}, \\
   \label{poisson-br2}
   && \{p_\mu,  f(y) \}_\theta = \frac{\partial f(y)}{\partial y^\mu} = \partial_\mu f(y), \\
   \label{poisson-br3}
   && \{\phi_\mu (y),  \phi_\nu (y) \}_\theta = -B_{\mu\nu} + \partial_\mu a_\nu(y) - \partial_\nu a_\mu (y)
    + \{a_\mu (y),  a_\nu (y) \}_\theta \equiv -B_{\mu\nu} + f_{\mu\nu} (y), \qquad
\end{eqnarray}
for a smooth function $f(y) \in C^\infty (N)$ and ``covariant momenta" $\phi_\mu (y),  \phi_\nu (y)
\in C^\infty (N)$ in Eq. \eq{x-phi}.
Therefore we have derived a Poisson algebra $\mathfrak{P} = (C^\infty (N), \{-, -\}_\theta)$
associated with a K\"ahler geometry we have started with.
Since we have introduced the equivalence relation \eq{eq-rel}, the coordinate transformations
generated by two vector fields in \eq{eq-rel} must be on the same gauge orbit, i.e.,
$\phi_\mu \sim \phi'_\mu = \phi_\mu + \delta_\lambda \phi_\mu$, where
$\delta_\lambda \phi_\mu = X_\lambda (\phi_\mu) = \{\phi_\mu (y),  \lambda (y) \}_\theta
= \partial_\mu \lambda (y) + \{ a_\mu (y),  \lambda (y) \}_\theta \equiv D_\mu \lambda (y)$.
This implies that the covariant momenta $\phi_\mu (y) \in C^\infty (U_\alpha)$ must be regarded as
local sections of a line bundle and $a_\mu (y)$ in the Darboux transformation \eq{x-phi} has to be regarded
as a gauge field whose field strength is given by $f_{\mu\nu} (y) =  \partial_\mu a_\nu(y)
- \partial_\nu a_\mu (y) + \{a_\mu (y),  a_\nu (y) \}_\theta$ in Eq. \eq{poisson-br3}.
Since they respect the non-Abelian structure due to the underlying Poisson structure \eq{p-bracket},
they are different from ordinary $U(1)$ gauge fields $A_\mu (x)$ in Eq. \eq{local-one-form},
so they will be called ``symplectic" $U(1)$ gauge fields.
One may notice that the Jacobi identity for the Poisson algebra $\mathfrak{P}$ leads to
the Bianchi identity for the symplectic $U(1)$ gauge fields:
\begin{equation}\label{symp-bianchi}
\{ \phi_a,  \{ \phi_b, \phi_c \}_\theta  \}_\theta
+ \{ \phi_b,  \{ \phi_c, \phi_a \}_\theta  \}_\theta
+ \{ \phi_c,  \{ \phi_a, \phi_b \}_\theta  \}_\theta
= D_a f_{bc} + D_b f_{ca} + D_c f_{ab} = 0.
\end{equation}

Since both sides of Eq. \eq{darboux-tr} are invertible, one can take its inverse and
derive the following relation
\begin{equation}\label{sw-map}
    \Theta^{ab} (x) \equiv \left(\frac{1}{B + F (x) } \right)^{ab}
    = \theta^{ac} \theta^{bd} \{\phi_c (y),  \phi_d (y)\}_\theta = \Big(\theta \big(B-f(y) \big) \theta \Big)^{ab}.
\end{equation}
From the above relation, it is easy to derive the exact Seiberg-Witten map between
commutative $U(1)$ gauge fields and symplectic $U(1)$ gauge fields \cite{nc-sw,esw-hsy}:
\begin{equation}\label{esw-map}
    f_{ab} (y) =  \left(\frac{1}{1 + F \theta} F \right)_{ab} (x).
\end{equation}
As was noticed before, the symplectic gauge fields are intrinsically non-Abelian
as well as non-linear as definitely indicated in Eq. \eq{esw-map}.
Therefore one can consider linear algebraic relations of symplectic $U(1)$ gauge fields
as a higher-dimensional analogue of four-dimensional self-duality equations such that
the equations of motion automatically follow, although we have not specified the underlying action yet.
Let us assume that they take the following form \cite{hsy-eujc,hsy-cy}
\begin{equation}\label{high-inst}
    \frac{1}{2} T_{abcd} f_{cd} = f_{ab}
\end{equation}
with a constant four-form tensor $T_{abcd}$. Using the Bianchi identity \eq{symp-bianchi},
it is easy to derive the ``equations of motion"
\begin{equation}\label{symp-eom}
 D_b f_{ab} = 0
\end{equation}
from the instanton equation \eq{high-inst}. Since the first-order partial differential equations \eq{high-inst}
are non-linear, it is reasonable to expect that there is a nontrivial regular solution satisfying them.
Such a solution, if any, will be called a symplectic $U(1)$ instanton.

Let us make it clear from the gravity perspective why symplectic $U(1)$ instantons should exist.
Recall that we have started with the K\"ahler geometry and have derived a natural Poisson algebra
associated with the K\"ahler geometry. Hence symplectic $U(1)$ gauge fields in the Poisson algebra
are the incarnation of K\"ahler geometry. From gravity point of view,
the generalized self-duality equation \eq{high-inst} imposes an additional condition on K\"ahler manifolds.
Therefore the symplectic $U(1)$ instantons should describe a particular class of K\"ahler manifolds.
A natural candidate is Calabi-Yau manifolds which are K\"ahler manifolds
of vanishing Ricci tensors \cite{besse}.
Recall that the Ricci tensor of a K\"ahler manifold takes an extremely simple form given by
\begin{equation}\label{ricci-kahler}
    R_{i\bar{j}} = - \frac{\partial^2 \ln \det g_{k\bar{l}}}{\partial z^i \partial \overline{z}^{\bar{j}}}.
\end{equation}
Therefore the Einstein equation for a K\"ahler metric reads
\begin{equation}\label{einstein-kahler}
   \partial_i  \partial_{\bar{j}} \ln \det g_{k\bar{l}} = 0.
\end{equation}
Local complex coordinates can be arranged in such a way that the Jacobians
$\det_{ij} (\partial z^i_\alpha/ \partial z^j_\beta)$ of the transition functions on $U_\alpha \bigcap U_\beta$
are one on all the overlaps. In that case $ \det g_{i\bar{j}}$ is a globally defined function and
the Einstein equation \eq{einstein-kahler} reduces to the Monge-Amp\`ere equation \cite{num-k3}
\begin{equation}\label{einstein-cy}
   \det g_{i\bar{j}} = \kappa,
\end{equation}
where the constant $\kappa$ is related to the volume of the K\"ahler manifold
that depends only on the K\"ahler class.
If so, one may guess that the generalized self-duality equation \eq{high-inst} is equivalent
to the Ricci-flat condition \eq{einstein-cy} of K\"ahler manifolds. It was proved in \cite{hea}
that the self-duality equation \eq{high-inst} is equivalent to the Einstein equation
\eq{einstein-cy} for $n = 2$ and $3$ cases. We speculate that it is true for any $n \geq 2$.
In four dimensions $(i.e. \; n=2)$, the Calabi-Yau manifolds are also called hyper-K\"ahler manifolds
or gravitational instantons. Thus the higher-dimensional Calabi-Yau manifolds may be regarded as
higher-dimensional gravitational instantons as Ricci-flat,  K\"ahler manifolds.
In the end, we have a remarkable equivalence between symplectic $U(1)$ instantons
and gravitational instantons \cite{gi-u1}.

Eq. \eq{esw-map} indicates that symplectic $U(1)$ instantons arise from the Seiberg-Witten
map of commutative $U(1)$ instantons which are singular in itself.
As was argued before, the symplectic $U(1)$ instantons
describe a smooth regular geometry without singularity because they are Calabi-Yau manifolds.
This is the reason why we need to allow singular $U(1)$ gauge fields to realize the K\"ahler gravity
in terms of a $U(1)$ gauge theory. We have seen that the gauge theory description of a K\"aher manifold
leads to symplectic $U(1)$ gauge fields rather than ordinary $U(1)$ gauge fields.\footnote{It should be the case
because the gravity is a non-linear theory. In order to make sense the equivalence between a K\"ahler gravity
and a gauge theory, the gauge symmetry structure of both theories should be compatible each other.
Symplectic $U(1)$ gauge fields respect a non-Abelian gauge symmetry whereas ordinary $U(1)$
gauge fields respect the Abelian gauge symmetry.}
To admit such a singular gauge field, the holomorphic line bundle $L$ is replaced by the rank one torsion
free sheaf which allows singular $U(1)$ gauge fields in real codimension four.
However the topological character of symplectic $U(1)$ gauge fields is obscure
from the gauge theory point of view. Although they are non-singular, their instanton
number--the second Chern class--is not quantized \cite{nc-sw}.
It may not be surprising since symplectic $U(1)$ gauge fields are sections of a line bundle
rather than its connections. They encode only the local geometry of a K\"ahler manifold and
need to be glued to encompass a global geometry.
Nevertheless their gravitational topology is well-defined since
the Euler characteristic and the Hirzebruch signature have integer numbers \cite{sly-prd}.
A nice feature is that, after quantization, NC $U(1)$ instantons \cite{nc-inst-ni} yield a well-defined topology
and their instanton number becomes an integer \cite{sakoetal}.
Accordingly the topology of quantized K\"ahler manifolds would be determined
by the topology of NC $U(1)$ gauge fields.

Let us define the action for symplectic $U(1)$ gauge fields.
The most natural variables that appear in the gauge theory description of a K\"ahler geometry
are covariant coordinates $x^a (y)$ or covariant momenta $\phi_a (y)$ in Eq. \eq{x-phi}.
Therefore we take the action as the form
\begin{equation}\label{symp-action}
    S = \frac{1}{4 g_{YM}^2} \int d^{2n} y \{\phi_a, \phi_b \}_\theta^2,
\end{equation}
where $g_{YM} = g (2\pi)^{n/2} l_s^{n-2}$ is a gauge coupling constant in this theory.
Note that the Lagrangian, $4 g_{YM}^2 \mathcal{L} = \{\phi_a, \phi_b \}^2_\theta
= B_{ab}^2 - 2 B_{ab} f_{ab} + f_{ab}^2$,
contains a background part, $B_{ab}^2$, and a total derivative term, $-2 B_{ab} f_{ab}$.
If one considers localized fluctuations only, one may drop the background part and
the total derivative term and the action for local fluctuations may be written as
\begin{equation}\label{f^2-action}
    S = \frac{1}{4 g_{YM}^2} \int d^{2n} y f_{ab}^2.
\end{equation}
The equations of motion \eq{symp-eom} can be derived from this action (or the action \eq{symp-action}).
However we will see that symplectic $U(1)$ gauge fields as well as NC $U(1)$ gauge fields
after quantization are not necessarily localized due to a subtle UV/IR mixing
in NC space and some fluctuations can be extended to macroscopic scales \cite{hsy-jpcs12}.
In this case the crossing term, $-2 B_{ab} f_{ab}$, cannot be dropped and it has
an important effect even at macroscopic scales. Moreover the background part, $B_{ab}^2$,
will have a surprising interpretation from the emergent gravity point of view.

\section{Quantization of K\"ahler Geometry and Noncommutative $U(1)$ Gauge Theory}

In the previous section we have shown that the K\"ahler geometry can be described by a $U(1)$ gauge theory
on a symplectic manifold with a slight generalization.
The gauge theory description of K\"ahler gravity is realized by viewing a K\"ahler manifold as
a phase space and its K\"ahler form as the symplectic two-form \cite{inov}.
This viewpoint naturally leads to a Poisson algebra $\mathfrak{P}$ associated
with the K\"ahler geometry we have started with. The Poisson algebra $\mathfrak{P}$ defines
an underlying algebraic structure of symplectic $U(1)$ gauge fields which
are local holomorphic sections of a line bundle.
The symplectic $U(1)$ gauge fields are the incarnation of K\"ahler geometry and correspond to
dynamical coordinates describing the deformation of a background symplectic structure.
Therefore a dynamical system defined by symplectic $U(1)$ gauge fields is described by
the Poisson algebra $\mathfrak{P}$.

Since the Poisson algebra $\mathfrak{P}$ defined by the Poisson bracket \eq{p-bracket} is mathematically
the same as the one in Hamiltonian dynamics of particles, one may try to quantize the Poisson algebra
in the exactly same way as quantum mechanics. Hence we apply the canonical quantization
to the Poisson algebra $\mathfrak{P} = (C^\infty (N), \{-, -\}_\theta)$ \cite{dirac}.
The canonical or Dirac quantization of $\mathfrak{P}$ consists of a suitable complex Hilbert space
$\mathcal{H}$ and a quantization map $\mathcal{Q}: C^\infty (N) \to \mathcal{A}_\theta$
from a commutative algebra $C^\infty (N)$ to a NC algebra $\mathcal{A}_\theta$.
The $\mathcal{Q}$-map associates to a function $f \in C^\infty (N)$ on $N$
a quantum operator $\widehat{f} \in \mathcal{A}_\theta$ acting on $\mathcal{H}$
by $f \mapsto \mathcal{Q} (f) \equiv \widehat{f}$. It should be $\mathbb{C}$-linear and
an algebra homomorphism:
\begin{equation}\label{q-prod}
    f \cdot g \mapsto \widehat{f*g} = \widehat{f} \cdot \widehat{g}
\end{equation}
and
\begin{equation}\label{*-product}
    f*g \equiv \mathcal{Q}^{-1} \big( \mathcal{Q} (f) \cdot \mathcal{Q} (g) \big)
\end{equation}
for $f, \, g \in C^\infty (N)$ and $\widehat{f}, \, \widehat{g} \in \mathcal{A}_\theta$.
Then the Poisson bracket \eq{p-bracket} in $\mathfrak{P}$ is mapped to a quantum bracket defined by
\begin{equation}\label{q-bracket}
   i \{f, g\}_\theta \mapsto \mathcal{Q} \big( [f,g]_* \big) = [\widehat{f}, \widehat{g}]
    = (\widehat{f} \cdot \widehat{g} - \widehat{g} \cdot \widehat{f} ),
\end{equation}
where
\begin{equation}\label{star-comm}
    [f, g]_* = f * g - g * f.
\end{equation}
Therefore the Poisson bracket controls the failure of commutativity
\begin{equation}\label{nc-poisson}
[\widehat{f}, \widehat{g}] \sim  i \{f, g\}_\theta + \mathcal{O} (\theta^2),
\end{equation}
so the Poisson bracket of classical observables may be seen as a shadow of the noncommutativity
in quantum world.

We now apply the quantization to the Poisson algebra $\mathfrak{P} = (C^\infty (N), \{-, -\}_\theta)$
of symplectic $U(1)$ gauge fields:
\begin{eqnarray}\label{q-gauge1}
    && \phi_a (y) \in C^\infty (N) \mapsto \mathcal{Q} (\phi_a) \equiv \widehat{\phi}_a (y)
    = p_a + \widehat{A}_a (y) \in \mathcal{A}_\theta, \\
    \label{q-gauge2}
    &&  \{\phi_a, \phi_b \}_\theta \in C^\infty (N) \mapsto - i \mathcal{Q}
    \big( [\phi_a, \phi_b]_* \big)
    = -i [ \widehat{\phi}_a (y), \widehat{\phi}_b (y)] =
    - B_{ab} + \widehat{F}_{ab} (y),
\end{eqnarray}
where $\widehat{A}_a = \mathcal{Q} (a_a)$ are called NC $U(1)$ gauge fields and
$\widehat{F}_{ab} = \mathcal{Q} (f_{ab})$ are their field strengths defined by
\begin{equation}\label{nc-fieldst}
    \widehat{F}_{ab} (y) = \partial_a \widehat{A}_b (y) - \partial_b \widehat{A}_a (y)
    - i [\widehat{A}_a (y), \widehat{A}_b (y)] \in \mathcal{A}_\theta.
\end{equation}
In particular, the background fields $y^\mu$ and $p_\mu$ satisfy the Heisenberg algebra
\begin{equation}\label{nc-r2n}
    [y^\mu, y^\nu] = i \theta^{\mu\nu}, \qquad [p_\mu, p_\nu] = - i B_{\mu\nu},
\end{equation}
where we have omitted the hat symbol for notational simplicity.
The NC algebra generated by the background fields will be denoted by $\mathbb{R}_\theta^{2n}$
and called the ``NC space" although the usual concept of space is not well-defined any more.
We see that NC $U(1)$ gauge fields act as the deformation of the ``vacuum" algebra $\mathbb{R}_\theta^{2n}$,
so they describe a dynamical NC space as Eq. \eq{q-gauge2} clearly shows.

Any NC algebra admits a nontrivial inner automorphism $\mathrm{Inn}(\mathcal{A}_\theta)$ defined by
\begin{equation}\label{inner}
\mathcal{O} \mapsto U \mathcal{O} U^{-1}
\end{equation}
for an operator $\mathcal{O} \in \mathcal{A}_\theta$ and $U \in \mathrm{Inn}(\mathcal{A}_\theta)$.
The infinitesimal generators of $\mathrm{Inn}(\mathcal{A}_\theta)$ act on $\mathcal{A}_\theta$ as
an inner derivation defined by
\begin{equation}\label{inn-der}
    \mathrm{ad}_{\widehat{f}} = -i [\widehat{f}, \, \cdot \,]
\end{equation}
given an operator $\widehat{f} \in \mathcal{A}_\theta$. The derivation obeys the Leibniz rule
\begin{equation}\label{leibniz}
 \mathrm{ad}_{\widehat{f}}(\mathcal{O}_1 \cdot \mathcal{O}_2) =
 \mathrm{ad}_{\widehat{f}}(\mathcal{O}_1) \cdot \mathcal{O}_2
 + \mathcal{O}_1 \cdot \mathrm{ad}_{\widehat{f}}(\mathcal{O}_2)
\end{equation}
for any two operators $\mathcal{O}_1, \mathcal{O}_2 \in \mathcal{A}_\theta$.
We denote the vector space of inner derivations as
\begin{equation}\label{der-d}
    \mathfrak{D} = \{ \mathrm{ad}_{\widehat{f}}: \widehat{g} \mapsto -i [\widehat{f}, \widehat{g}] |
    \widehat{f} \in \mathcal{A}_\theta \}.
\end{equation}
For example, the background operators $p_\mu$ act
as a differential operator, i.e.,
\begin{equation}\label{inn-der}
    \mathrm{ad}_{p_\mu} = \partial_\mu
\end{equation}
and the covariant momenta $\widehat{\phi}_a \in \mathcal{A}_\theta$ act as a covariant derivative given by
\begin{equation}\label{co-moment}
     \mathrm{ad}_{\widehat{\phi}_a} \mathcal{O} = -i [\widehat{\phi}_a, \mathcal{O}]
     =\partial_a \mathcal{O} - i [\widehat{A}_a, \mathcal{O}] \equiv \widehat{D}_a \mathcal{O}
\end{equation}
for an observable $\mathcal{O}  \in \mathcal{A}_\theta$. For the latter case, the finite inner
automorphism in \eq{inner} is given by $U_c = \exp \big(-i c^a \widehat{\phi}_a \big)
\in \mathrm{Inn}(\mathcal{A}_\theta)$ with $c^a \in \mathbb{R}$ or $\mathbb{C}$.
Since the inner derivations $\mathrm{ad}_{\widehat{\phi}_a}$ are important operators
for our later application, we denote them by \cite{q-emg,hsy-inflation}
\begin{equation}\label{nc-vector}
 \mathrm{ad}_{\widehat{\phi}_a} \equiv \widehat{V}_a \in \mathfrak{D}.
\end{equation}
The linear map $\rho: \mathcal{A}_\theta \to \mathfrak{D}$ is a Lie algebra homomorphism
because it satisfies the relation
\begin{equation}\label{lie-homo}
   \mathrm{ad}_{[\widehat{f}, \widehat{g}]} = i [ \mathrm{ad}_{\widehat{f}}, \mathrm{ad}_{\widehat{g}}]
\end{equation}
for any $\widehat{f}, \widehat{g} \in \mathcal{A}_\theta$. One can easily check Eq. \eq{lie-homo}
using the Jacobi identity of the NC algebra $\mathcal{A}_\theta$.

Denote the center of $\mathcal{A}_\theta$ by $\mathcal{Z}(\mathcal{A}_\theta) = \{\widehat{f}
\in \mathcal{A}_\theta| [\widehat{f}, \widehat{g}] = 0, \forall \widehat{g} \in \mathcal{A}_\theta \}$
and introduce an equivalence relation $\widehat{f} \sim \widehat{g}$ in $\mathcal{A}_\theta$
if and only if $\widehat{g} = \widehat{f} + c$ with $c \in \mathcal{Z}(\mathcal{A}_\theta)$.
Consider the set of equivalence classes of $\mathcal{A}_\theta$ by $\sim$,
denoted as $\widetilde{\mathcal{A}}_\theta = \mathcal{A}_\theta/\sim$.
Then the linear map $\widetilde{\rho}: \widetilde{\mathcal{A}}_\theta \to \mathfrak{D}$ is
the Lie algebra isomorphism. One important example is
\begin{equation}\label{imp-exam}
-i \, \mathrm{ad}_{[\widehat{\phi}_a, \widehat{\phi}_b]} = \mathrm{ad}_{\widehat{F}_{ab}}
=  [\mathrm{ad}_{\widehat{\phi}_a}, \mathrm{ad}_{\widehat{\phi}_b}] = [\widehat{V}_a, \widehat{V}_b]
\end{equation}
where we used the fact that $-i [ \widehat{\phi}_a, \widehat{\phi}_b] = - B_{ab} + \widehat{F}_{ab}$
and $B_{ab} \in \mathcal{Z}(\mathcal{A}_\theta)$. The fact that the derivation $\mathfrak{D}$ is inert
for elements in $\mathcal{Z}(\mathcal{A}_\theta)$ is the crux to resolve the cosmological constant problem
in general relativity \cite{hsy-jhep09}, as will be discussed later.

After the quantization $\mathcal{Q}: C^\infty (N) \to \mathcal{A}_\theta$,
the classical action \eq{symp-action} is lifted to the action of NC $U(1)$ gauge fields given by
\begin{equation}\label{q-action}
    \widehat{S} = - \frac{1}{4 g_{YM}^2} \int d^{2n} y [\widehat{\phi}_a, \widehat{\phi}_b ]^2,
\end{equation}
where the integral is defined as the trace over a separable Hilbert space $\mathcal{H}$
on which the operators $\widehat{\phi}_a$ act, i.e.,
\begin{equation}\label{int-tr}
    \int \frac{d^{2n} y}{(2\pi)^n |\mathrm{Pf} (\theta)|} = \mathrm{Tr}_{\mathcal{H}}.
\end{equation}
We will assume that $\mathrm{Tr}_{\mathcal{H}} [\widehat{f}, \widehat{g}] = 0$ if at least one of
the operators $(\widehat{f}, \widehat{g})$ is compactly supported.
The equations of motion derived from \eq{q-action} read as
\begin{equation}\label{nc-eom}
    \widehat{D}_b \widehat{F}_{ab} = 0.
\end{equation}
Similarly, the generalized self-duality equation \eq{high-inst} is now defined as an
operator algebra\footnote{The four-form tensor $T_{abcd}$ in the self-duality equation \eq{high-ncinst}
breaks the rotational symmetry to a subgroup $H \subset O(2n)$. Thus the two-form field strength
$\widehat{F} = \frac{1}{2} \widehat{F}_{ab} dy^a \wedge dy^b$ can be classified by the unbroken symmetry $H$
under which $T_{abcd}$ remain invariant. The self-duality equation \eq{high-ncinst} projects the two-form
$\widehat{F}$ to an invariant subspace.}
\begin{equation}\label{high-ncinst}
    \frac{1}{2} T_{abcd} \widehat{F}_{cd} = \widehat{F}_{ab}.
\end{equation}
In four and six dimensions, the above self-duality equation is given by
\begin{equation} \label{t-tensor}
 T_{abcd} = \left\{
  \begin{array}{ll}
   \pm \varepsilon_{abcd}, & \hbox{$n=2$;} \\
   - \frac{1}{2}\varepsilon_{abcdef} \varepsilon^{ef}, & \hbox{$n=3$,}
  \end{array}
\right.
\end{equation}
where $\varepsilon$'s are volume tensors in each dimension and $\varepsilon^{ef} = \frac{\theta^{ef}}{\alpha'}$.
The self-duality equation \eq{high-ncinst} admits nontrivial regular solutions
called NC $U(1)$ instantons \cite{nc-inst-ni} and NC Hermitian $U(1)$ instantons \cite{nonins,hsy-cy}
in four and six dimensions, respectively.

Let us understand what Eqs. \eq{nc-eom} and \eq{high-ncinst} mean from the K\"ahler gravity point of view.
Our journey so far may be summarized by the following diagram:
\begin{equation} \label{diag}
\begin{array}[c]{ccc}
\mathrm{K\ddot{a}hler~gravity}&\stackrel{\mathfrak{I}^{-1}_\epsilon}{\longrightarrow}&
\mathrm{Symplectic~{\it U(1)}~gauge~theory }\\
{\mathcal{Q}}\downarrow\scriptstyle&&\downarrow{\mathcal{Q}}\scriptstyle\\
? &\stackrel{\mathfrak{I}_\theta}{\longleftarrow}&\mathrm{NC~{\it U(1)}~gauge~theory }
\end{array}
\end{equation}
Here $\mathcal{Q}: C^\infty (N) \to \mathcal{A}_\theta$ means the quantization we have defined
at the beginning of this section and $\mathfrak{I}$ means an isomorphism between two theories.
In some sense $\mathfrak{I}$ corresponds to the gauge-gravity duality. We will see that
it can be interpreted as the large $N$ duality too. Since symplectic $U(1)$ gauge theory is a commutative limit of
NC $U(1)$ gauge theory in the sense of Eq. \eq{nc-poisson}, we understand the classical isomorphism in \eq{diag}
as $\mathfrak{I}_\epsilon = \mathfrak{I}_\theta|_{\varepsilon = |\theta| \to 0}$.
According to the flow chart in \eq{diag}, it is reasonable to identify the unknown theory
with a quantized K\"ahler gravity. Actually this relation was already observed in \cite{inov}
in the context of topological strings probing K\"ahler manifolds where
several nontrivial evidences have been analyzed to support the picture.
In particular, the authors in \cite{inov} argue that
NC $U(1)$ gauge theory is the fundamental description of K\"ahler gravity at all scales including
the Planck scale and provides a quantum gravity description such as quantum gravitational foams.
The duality in \cite{inov,mnop} has been further clarified in \cite{neova-kap} by showing that it follows
from the S-duality of the type IIB superstring. So we claim the following duality:
\begin{equation} \label{q-diag}
\begin{array}[c]{ccc}
\mathrm{K\ddot{a}hler~gravity}&\stackrel{\mathfrak{I}^{-1}_\epsilon}{\longrightarrow}&
\mathrm{Symplectic~{\it U(1)}~gauge~theory }\\
{\mathcal{Q}}\downarrow\scriptstyle&&\downarrow{\mathcal{Q}}\scriptstyle\\
\mathrm{Quantized~K\ddot{a}hler~gravity} &\stackrel{\mathfrak{I}_\theta}{\longleftarrow}&
\mathrm{NC~{\it U(1)}~gauge~theory }
\end{array}
\end{equation}
This duality, if any, suggests an important clue about how to quantize the K\"ahler gravity.
Surprisingly, the correct variables for quantization are not metric fields but dynamical coordinates
$x^a(y)$ or $\phi_a(y) = B_{ab} x^b (y)$ and their quantization is defined in terms of $\alpha'$
rather than $\hbar$. So far, there is no well-established clue to quantize metric fields directly
in terms of $\hbar$ in spite of impressive developments in loop quantum gravity \cite{lqg}.
However, the picture in \eq{q-diag} suggests a completely new quantization scheme \cite{q-emg}
where quantum gravity is defined by quantizing spacetime itself in terms of $\alpha'$,
leading to a dynamical NC spacetime described by a NC $U(1)$ gauge theory.

The duality relation in \eq{q-diag} may be more accessible with the corresponding relation for solutions
of the self-duality equation \eq{high-ncinst}. It means that the duality relation \eq{q-diag} is
restricted to a particular class of K\"ahler manifolds with vanishing Ricci tensors as
we have discussed in section 2:
\begin{equation} \label{cy-diag}
\begin{array}[c]{ccc}
\mathrm{Calabi-Yau~manifold}&\stackrel{\mathfrak{I}^{-1}_\epsilon}{\longrightarrow}&
\mathrm{Symplectic~{\it U(1)}~instanton }\\
{\mathcal{Q}}\downarrow\scriptstyle&&\downarrow{\mathcal{Q}}\scriptstyle\\
\mathrm{Quantized~Calabi-Yau~manifold} &\stackrel{\mathfrak{I}_\theta}{\longleftarrow}&
\mathrm{NC~{\it U(1)}~instanton }
\end{array}
\end{equation}
It was shown in \cite{hsy-cy,hsy-mirror} that symplectic $U(1)$ instantons obeying
the self-duality equation \eq{high-inst} are equivalent to Calabi-Yau manifolds.
Since Eq. \eq{high-inst} simply arises at the commutative limit of the NC self-duality
equation \eq{high-ncinst}, it was claimed in \cite{hsy-cy} that the duality relation $\mathfrak{I}_\theta$
should be true in quantum level too as depicted in \eq{cy-diag}.
The duality relation \eq{q-diag} in six dimensions has been illuminated in \cite{inov}
by showing the complete equivalence of the topological vertex counting of
the all-genus string partition function with the partition function of $U(1)$ maximally
supersymmetric topologically twisted gauge theory on toric Calabi-Yau manifolds.

An important basis for the furtive isomorphism $\mathfrak{I}_\epsilon$ in \eq{q-diag} is that
the $U(1)$ gauge theory also respects the equivalence principle in K\"ahler gravity.
So it should be interesting to see how the equivalence principle in K\"ahler gravity is lifted
to a quantum world, dubbed {\it quantum equivalence principle} \cite{hsy-review,q-emg} in quantized K\"ahler gravity.
First let us recapitulate an underlying logic in the duality relation \eq{q-diag}.
Since a K\"ahler manifold is a symplectic manifold too, one can apply the Darboux theorem
to the local K\"ahler form \eq{kahler-gauge} which is also regarded as a symplectic two-form
on a local patch $U_\alpha$. It reads as $\varphi_\alpha^* (B + F_\alpha) = B$ with
a local coordinate transformation $\varphi_\alpha \in \mathrm{Diff}(U_\alpha)$.
Since the local K\"ahler form $\mathcal{F}_\alpha = B + F_\alpha$ on $U_\alpha$ is
isomorphically mapped to a local K\"ahler metric given a fixed complex structure,
the Darboux transformation can be equally stated in terms of the metric as
$\varphi_\alpha^* (g_{i\bar{j}}) = \delta_{i\bar{j}}$.\footnote{It should be
understood with a caveat as we explained in the paragraph below Eq. \eq{darboux-tr}.}
Thus the fact that the Darboux transformation locally eliminates dynamical $U(1)$ gauge fields is equally phrased
as a local trivialization of an underlying K\"ahler manifold according to the equivalence principle
in general relativity. The NC $U(1)$ gauge theory is constructed by lifting the coordinate
transformation \eq{x-phi} to a local automorphism of $\mathcal{A}_\theta$ defined by
$\mathcal{D}_A = \mathcal{Q} \circ \varphi^*_\alpha$ which acts on the coordinates $y^a$ as
\begin{equation}\label{auto-q}
  \mathcal{D}_A (y^a) = \widehat{x}^a (y) = y^a + \theta^{ab} \widehat{A}_b (y)
= \theta^{ab} \widehat{\phi}_b (y) \in \mathcal{A}_\theta.
\end{equation}
They obey the commutation relations
\begin{equation}\label{q-comm}
    [\widehat{x}^a, \widehat{x}^b] = i \big( \theta - \theta\widehat{F}\theta \big)^{ab} (y)
\equiv  i \widehat{\Theta}^{ab} (y), \qquad
[\widehat{\phi}_a, \widehat{\phi}_b] = - i \big( B_{ab} - \widehat{F}_{ab} (y) \big).
\end{equation}

Now let us take the bivector $\Theta = \frac{1}{2} \Theta^{ab} (x) \frac{\partial}{\partial x^a} \bigwedge
\frac{\partial}{\partial x^b} \in \Gamma(\Lambda^2 TN)$ given by Eq. \eq{sw-map}.
Since $d\mathcal{F}=0$ with $\mathcal{F} = B + F$, the Schouten-Nijenhuis bracket of
the bivector $\Theta \in \Gamma(\Lambda^2 TN)$ identically vanishes, that is $[\Theta, \Theta]_{SN} = 0$.
Thus the bivector $\Theta$ defines a new Poisson structure on $N$.
The commutative algebra $C^\infty (N)$ forms a Lie algebra $\mathfrak{P}'
= \big(C^\infty (N), \{ -, - \}_\Theta \big)$ under the Poisson bracket
\begin{equation}\label{new-poisson}
    \{ f, g \}_\Theta = \Theta (df, dg)  \qquad \mathrm{or} \qquad  \{f, g\}_\Theta
= \Theta^{\mu\nu} (x) \frac{\partial f(x)}{\partial x^\mu}\frac{\partial g(x)}{\partial x^\nu}.
\end{equation}
The quantization map associates to a function $f \in C^\infty (N)$ on $N$
a quantum operator $\widetilde{f} \in \mathcal{A}_{\widehat{\Theta}}$ acting on $\mathcal{H}$
by $f \mapsto \mathcal{D}_A (f) \equiv \widetilde{f}$. It should also be $\mathbb{C}$-linear and
an algebra homomorphism:
\begin{equation}\label{q'-prod}
    f \cdot g \mapsto \widetilde{f*'g} = \widetilde{f} \cdot \widetilde{g}
\end{equation}
and
\begin{equation}\label{*'-product}
    f*'g \equiv \mathcal{D}_A^{-1} \big( \mathcal{D}_A (f) \cdot \mathcal{D}_A (g) \big)
\end{equation}
for $f, \, g \in C^\infty (N)$ and $\widetilde{f}, \, \widetilde{g} \in \mathcal{A}_{\widehat{\Theta}}$.
Then the Poisson bracket \eq{new-poisson} in $\mathfrak{P}'$ is mapped to a quantum bracket defined by
\begin{equation}\label{q'-bracket}
   i \{f, g\}_\Theta \mapsto \mathcal{D}_A \big( [f,g]_{*'} \big) = [\widetilde{f}, \widetilde{g}]
    = (\widetilde{f} \cdot \widetilde{g} - \widetilde{g} \cdot \widetilde{f} ),
\end{equation}
where
\begin{equation}\label{star'-comm}
    [f, g]_{*'} = f *' g - g *' f.
\end{equation}
Therefore the Poisson algebra $\mathfrak{P}'$ arises as the commutative limit of
the NC algebra $\mathcal{A}_{\widehat{\Theta}}$, i.e.,
\begin{equation}\label{nc'-poisson}
[\widetilde{f}, \widetilde{g}] \sim  i \{f, g\}_\Theta + \mathcal{O} (\Theta^2).
\end{equation}
The new Poisson bracket \eq{new-poisson} is related to the old one \eq{p-bracket} by
a Poisson map \cite{q-emg}
\begin{equation}\label{poisson-map}
    \varphi_\alpha^* \{f, g\}_\Theta = \{ \varphi_\alpha^* f,  \varphi_\alpha^* g\}_\theta
\end{equation}
for all $f, g \in C^\infty (N)$.

One can see from Eq. \eq{q'-bracket} that $[\widetilde{x}^a, \widetilde{x}^b]
= \mathcal{D}_A \big( [x^a, x^b]_{*'} \big) = i \mathcal{D}_A \big( \Theta^{ab} (x) \big)
= i \widetilde{\Theta}^{ab} (\widetilde{x})$. Thus the commutation relations \eq{q-comm}
can be derived from the NC algebra $\mathcal{A}_{\widehat{\Theta}}$ with the identification
$\widetilde{x}^a := \widehat{x}^a (y)$ and $\widetilde{\Theta}^{ab} (\widetilde{x}) :=
\widehat{\Theta}^{ab} (y)$. This means that the new NC algebra $\mathcal{A}_{\widehat{\Theta}}$
can be related to the old one $\mathcal{A}_{\theta}$ by choosing the quantization
$\mathcal{D}_A = \mathcal{Q} \circ \varphi^*_\alpha$. Since the local diffeomorphism
$\varphi_\alpha$ on $U_\alpha$ is a coordinate transformation to a Darboux frame defined by
$\varphi_\alpha^* (B + F_\alpha) = B$ or a locally inertial frame given by
$\varphi_\alpha^* (\delta + h) = \delta$, the quantization $\mathcal{D}_A = \mathcal{Q} \circ \varphi^*_\alpha$
corresponds to a lift of the classical Darboux theorem or the equivalence principle to quantized
NC algebras. Indeed it was shown in \cite{morita} that two star products $*$ and $*'$
are Morita equivalent if and only if they are, modulo diffeomorphisms, related by the action of the Picard group
${\tt Pic} (C^\infty(N)) \cong H^2 (N, \mathbb{Z})$ given by an element of isomorphism classes of
holomorphic line bundles over $N$. In our case this is true for the $\Lambda$-symmetry \eq{b-transf}.
Naturally the NC algebras $\mathcal{A}_{\theta}$ and $\mathcal{A}_{\widehat{\Theta}}$ are also
Morita equivalent according to the invertible covariance map \eq{*'-product}.
So one may consider the Morita equivalence between $\mathcal{A}_{\theta}$ and $\mathcal{A}_{\widehat{\Theta}}$
as the {\it quantum equivalence principle} \cite{q-emg}.

Given a Poisson matrix as the form \eq{symp-theta} for the NC space $\mathbb{\mathbb{R}}^{2n}$,
the Heisenberg algebra \eq{nc-r2n} can be written as
\begin{equation}\label{ho-algebra}
    [a_i, a_j^\dagger] = \delta_{ij},  \qquad i,j = 1, \cdots, n,
\end{equation}
where $a_i = \frac{y^{2i-1} + \sqrt{-1}y^{2i}}{\sqrt{2\alpha'}}$ and $a_i^\dagger = \frac{y^{2i-1} - \sqrt{-1}y^{2i}}{\sqrt{2\alpha'}}$. The Hilbert space for the representation of the algebra \eq{ho-algebra}
is given by the Fock space
\begin{equation}\label{fock-space}
    \mathcal{H} = \{ | \mathbf{n} \rangle \equiv |n_1, n_2, \cdots, n_n \rangle \, |
n_1, n_2, \cdots, n_n = 0, 1, \cdots, \infty \}.
\end{equation}
The basis of the Fock space is orthonormal, i.e., $\langle\mathbf{n}|\mathbf{m} \rangle
= \delta_{\mathbf{n},\mathbf{m}}$ and
complete, i.e., $\sum_{\mathbf{n} = 0}^{\infty} | \mathbf{n}\rangle \langle \mathbf{n}| = \mathbf{1}_{\mathcal{H}}$,
as is well-known from quantum mechanics. Since the Fock space (\ref{fock-space}) has a countable basis,
it is convenient to introduce a one-dimensional basis using the ``Cantor diagonal method" to put the
$n$-dimensional non-negative integer lattice in $\mathcal{H}$ into one-to-one correspondence
with the natural numbers:
\begin{equation}\label{cantor}
\mathbb{Z}^n_{\geq 0} \leftrightarrow \mathbb{N}: |\mathbf{n}\rangle \leftrightarrow |n \rangle, \;
n = 1, \cdots, N \to \infty.
\end{equation}
In this one-dimensional basis, the completeness relation of the Fock space (\ref{fock-space}) is now
given by $\sum_{n = 1}^{\infty} | n \rangle \langle n| = \mathbf{1}_{\mathcal{H}}$.

It is known \cite{dirac} that the representation of NC operators on the Fock space $\mathcal{H}$ is
given by $N \times N$ matrices where $N = \mathrm{dim}(\mathcal{H}) \to \infty$.
Consider two arbitrary dynamical fields $\widehat{f}(y)$ and $\widehat{g}(y)$ on the $2n$-dimensional
NC space $\mathbb{R}^{2n}_{\theta}$ which are elements of the NC algebra $\mathcal{A}_\theta$.
Since the dynamical variables in $\mathcal{A}_\theta$ can be regarded
as linear operators acting on the Hilbert space (\ref{fock-space}), we can represent the operators
as $N \times N$ matrices in $\mathrm{End}(\mathcal{H})
\equiv \mathcal{A}_N$ where $N = \mathrm{dim}(\mathcal{H}) \to \infty$:
\begin{eqnarray}	
\begin{array}{rcl}
     && \widehat{f}(y) = \sum_{n,m=1}^\infty | n \rangle \langle n| \widehat{f} (y) | m \rangle \langle m|
      := \sum_{n,m=1}^\infty F_{nm} | n \rangle \langle m|, \\
     && \widehat{g}(y) = \sum_{n,m=1}^\infty | n \rangle \langle n| \widehat{g} (y) | m \rangle \langle m|
      := \sum_{n,m=1}^\infty G_{nm} | n \rangle \langle m|,
\end{array}
\label{matrix-rep}
\end{eqnarray}
where $F$ and $G$ are $N \times N$ matrices in $\mathcal{A}_N = \mathrm{End}(\mathcal{H})$.
Then we get a natural composition rule
\begin{equation*}
 \widehat{f}(y) \cdot \widehat{g}(y)  = \sum_{n,l,m=1}^\infty | n \rangle \langle n|
\widehat{f} (y) | l \rangle \langle l| \widehat{g}(y) | m \rangle \langle m|
      = \sum_{n,l,m=1}^\infty F_{nl} G_{lm} | n \rangle \langle m|.
\end{equation*}
The above composition rule implies that the ordering in the NC algebra $\mathcal{A}_\theta$
is perfectly compatible with the ordering in the matrix algebra $\mathcal{A}_N$.
Thus we can straightforwardly translate multiplications of NC fields in $\mathcal{A}_\theta$
into those of matrices in $\mathcal{A}_N$ using the matrix representation (\ref{matrix-rep})
without any ordering ambiguity. Indeed the linear representation $\rho: \mathcal{A}_\theta \to
\mathcal{A}_N$ on the Hilbert space $\mathcal{H}$ is a Lie algebra homomorphism.
Using the map (\ref{matrix-rep}),
the trace over $\mathcal{A}_\theta$ can also be transformed into the trace over $\mathcal{A}_N$, i.e.,
\begin{equation}\label{matrix-trace}
    \int \frac{d^{2n} y}{(2 \pi )^n |\mathrm{Pf}\theta|} = \mathrm{Tr}_{\mathcal{H}} = \mathrm{Tr}_N.
\end{equation}
Then it is straightforward to map the NC $U(1)$ gauge theory \eq{q-action} to the matrix action \cite{ikkt}
\begin{equation}\label{matrix-action}
    S = - \frac{1}{4 g_0^2} \mathrm{Tr}_N [\Phi_a, \Phi_b]^2
\end{equation}
where $g_0 = \frac{g_{YM}}{\sqrt{(2\pi)^n |\mathrm{Pf}(\theta)|}} = \frac{g}{\alpha'}$
is the coupling constant of matrix model and $\Phi_a \in \mathcal{A}_N, \; a=1, \cdots, 2n,$ is
the matrix representation of a covariant momentum $\widehat{\phi}_a \in \mathcal{A}_\theta$.

Note that the matrix algebra $\mathcal{A}_N$ is
an associative algebra with product $AB$ and forms a Lie algebra under the bracket
$[A, B] = AB - BA$ for $A, B \in \mathcal{A}_N$.
Then the Jacobi identity
\begin{equation}\label{m-jacobi}
    [\Phi_a, [\Phi_b, \Phi_c]] + [\Phi_b, [\Phi_c, \Phi_a]] + [\Phi_c, [\Phi_a, \Phi_b]]= 0
\end{equation}
holds due to the associativity of the algebra $\mathcal{A}_N$.
The matrix model is further subject to the equations of motion
\begin{equation}\label{m-eom}
    [\Phi_b, [\Phi_a, \Phi_b]] = 0
\end{equation}
derived from the principle of least action.
It is well-known that every automorphism of the matrix algebra $\mathcal{A}_N$ is an inner automorphism.
More precisely, we have the following result. Let $\Psi: \mathcal{A}_N \to \mathcal{A}_N$ be a bijective
linear map satisfying $\Psi(AB) = \Psi(A) \Psi(B), \; A, B \in \mathcal{A}_N$. Then there exists an
invertible matrix $U \in \mathcal{A}_N$ such that
\begin{equation}\label{inner-auto}
    \Psi(A) = U A U^{-1}
\end{equation}
for every $A \in \mathcal{A}_N$. It will be called the $U(N)$ gauge symmetry from the matrix model viewpoint.
In our case we need to take the limit $N \to \infty$.
Since we want to achieve a background-independent formulation of quantum gravity in terms of the algebra
of (large $N$) matrices, this property will be important to define a deformation complex of $\mathcal{A}_N$.
The matrix action \eq{matrix-action} has an extra global automorphism given by
\begin{equation}\label{g-auto}
    \Phi_a \to R_{ab} \Phi_b + c_a
\end{equation}
where $(R)_{ab}$ is a rotation in $SO(2n)$ and $c_a$ are constants proportional to the identity matrix.

The final destination of our journey started with K\"ahler geometry is a zero-dimensional
matrix model defined by the action \eq{matrix-action}.
The matrix model has been derived from the noncommutative $U(1)$ gauge theory which
arguably describes a quantized K\"ahler geometry according to the duality picture \eq{q-diag}.
Thus the correspondence in \eq{q-diag} is actually a large $N$ duality between gravity
and matrix model.\footnote{\label{ads-cft}If the rank of $\theta^{\mu\nu}$ is less than $2n$,
say, $2m$ and $d = 2(n-m) \geq 0$, NC fields $\widehat{\phi} (x, y)$ are elements
of the NC algebra $\mathcal{A}_\theta^d \equiv \mathcal{A}_\theta \big( C^\infty (\mathbb{R}^d) \big)
= C^\infty (\mathbb{R}^d) \otimes \mathcal{A}_\theta$
where $x \in \mathbb{R}^d$ and $y \in \mathbb{R}_\theta^{2m}$. The matrix representation
$\rho: \mathcal{A}_\theta^d \to \mathcal{A}_N^d$ of $\widehat{\phi} (x, y)$ gives us a matrix
$\Phi (x) \in \mathcal{A}_N^d \equiv \mathcal{A}_N \big( C^\infty (\mathbb{R}^d) \big) = C^\infty (\mathbb{R}^d)
\otimes \mathcal{A}_N$ defined on the commutative space $\mathbb{R}^d$. In that case
we will get a $d$-dimensional $U(N \to \infty)$ gauge theory after the matrix representation \cite{q-emg,hsy-2016}.
For instance, ten-dimensional NC $U(1)$ gauge theory on $\mathbb{R}^4 \times \mathbb{R}_\theta^6$ (i.e., $n=5$ and $m=3$) leads to the four-dimensional $U(N \to \infty)$ gauge theory on $\mathbb{R}^4$ which is the bosonic part
of four-dimensional $\mathcal{N}=4$ supersymmetric gauge theory in the AdS/CFT correspondence.
The duality \eq{q-diag} in this case corresponds to the gauge-gravity duality on an asymptotically flat spacetime background.}
Since the matrix model \eq{matrix-action} is a zero-dimensional theory,
any concept of space (and time) is not necessary to define the theory.
Rather the concept of geometry in K\"ahler gravity has been replaced by NC algebras $\mathcal{A}_\theta$
and $\mathcal{A}_N$. In this sense we have arrived at a completely background-independent
formulation of quantized K\"ahler geometry.

\section{Matrix Model and Quantum Gravity}

We have started with K\"ahler geometry and have arrived at a radical endpoint.
The concept of geometry has been completely disappeared at the endpoint.
Now we want to play with the zero-dimensional matrix model \eq{matrix-action} to examine how to recover
our starting point--K\"ahler geometry--from the background-independent formulation.

First let us specify vacua of the matrix model known as the Coulomb branch.
The conventional choice of vacuum in the Coulomb branch of $U(N)$ Yang-Mills theory is given by
\begin{equation}\label{c-coulomb}
  [\Phi_a, \Phi_b]|_{\mathrm{vac}} = 0 \qquad \Longrightarrow \qquad
\langle \Phi_a \rangle_{\mathrm{vac}} = \mathrm{diag} \big((\upsilon_a)_1, (\upsilon_a)_2, \cdots, (\upsilon_a)_N \big),
\end{equation}
where $(\upsilon_a)_i \; (a = 1, \cdots, 2n, \; i=1, \cdots, N)$ are constant vacuum expectation values, possibly zeros.
Depending on a specific symmetry breaking pattern, the $U(N)$ gauge symmetry is reduced to
$G_{k_1} \times G_{k_2} \times \cdots \times G_{k_p}$ where $G_{k_i} = SU(k_i-1)$ or $U(k_i)$ and $\sum_{i=1}^p k_i =N$.
If $(\upsilon_a)_i$ are all different, the $U(N)$ gauge symmetry is broken to $U(1)^N$.
One can study a low energy effective action by the expansion $\Phi_a = \langle \Phi_a \rangle_{\mathrm{vac}}
+ \delta \Phi_a$ around the diagonal configurations \eq{c-coulomb} in the Coulomb branch.
However we fail to reproduce the $2n$-dimensional NC $U(1)$ gauge theory \eq{q-action} in that way.
It is not difficult to find a correct vacuum to derive a higher-dimensional NC $U(1)$ gauge theory from the matrix
model \eq{matrix-action}. In retrospect, it is required to have a separable Hilbert space to achieve the matrix representation $\rho: \mathcal{A}_\theta \to \mathcal{A}_N$. The Hilbert space \eq{fock-space} arises
as a linear representation of the vacuum algebra \eq{nc-r2n}.
Therefore it is obvious what is the correct vacuum for our purpose.
It is given by a coherent vacuum in the so-called NC Coulomb branch
defined by \cite{japan,q-emg}
\begin{equation}\label{nc-coulomb}
[\Phi_a, \Phi_b]|_{\mathrm{vac}} = - i B_{ab} \qquad \Longrightarrow \qquad
\langle \Phi_a \rangle_{\mathrm{vac}} = p_a = B_{ab} y^b.
\end{equation}
We have learned from quantum mechanics \cite{dirac} that such a NC Coulomb branch
is allowed when the size of matrices
is infinite. Since we are considering the limit $N \to \infty$, the large $N$ limit
opens a new phase of the Coulomb branch given by \eq{nc-coulomb}.
Unfortunately the NC Coulomb branch in gauge-gravity duality has been mostly ignored so far.
Note that the NC Coulomb branch \eq{nc-coulomb} saves the NC nature of matrices
while the conventional commutative vacuum \eq{c-coulomb} dismisses the property.

Suppose that the fluctuations around the vacuum \eq{nc-coulomb} take the form
\begin{equation}\label{nc-fluct}
    \Phi_a = p_a + \widehat{A}_a (y).
\end{equation}
By considering the matrices $\Phi_a \in \mathcal{A}_N$ as a linear representation of the operators
$\widehat{\phi}_a$ on the Hilbert space $\mathcal{H}$ as in Eq. \eq{matrix-rep},
one can associate the matrix algebra
$\mathcal{A}_N = \mathrm{End}(\mathcal{H})$ with a NC algebra $\mathcal{A}_\theta$.
Then it is straightforward to get the $2n$-dimensional NC $U(1)$ gauge theory \eq{q-action}
by plugging the expansion \eq{nc-fluct} into the matrix action \eq{matrix-action} and
using the relation \eq{matrix-trace}. It may be emphasized that the NC Coulomb branch \eq{nc-coulomb}
is a consistent vacuum since it obeys the equations of motion \eq{m-eom}.
It is also the crux to realize the equivalence between a large $N$ matrix model and
a higher-dimensional NC $U(1)$ gauge theory \cite{japan}.
If the conventional commutative vacuum \eq{c-coulomb} were chosen, we would have failed to
realize the equivalence. Indeed it turns out \cite{q-emg,hsy-2016,hsy-jhep09} that
the NC Coulomb branch is crucial to realize the emergent gravity from matrix models
or large $N$ gauge theories.

Since we have already obtained the $2n$-dimensional NC $U(1)$ gauge theory \eq{q-action}
from the zero-dimensional matrix model \eq{matrix-action} by considering fluctuations
around the NC Coulomb branch \eq{nc-coulomb}, it would be affirmative to yield
the K\"ahler geometry again by reversing the procedure for the flowchart of the duality \eq{q-diag}.
However it would be desirable to develop a more systematic way to derive the classical geometry from
a matrix model and a NC $U(1)$ gauge theory.
Before doing that, let us first check where the flat space $\mathbb{R}^{2n}$ comes from.
Definitely the flat space corresponds to a global K\"aher form $\mathcal{F} = B$ in Eq. \eq{kahler-gauge}
without any local fluctuations. From the matrix model perspective, it arises from the vacuum
in the NC Coulomb branch \eq{nc-coulomb} whose density can be evaluated
by the action \eq{q-action}:
\begin{equation}\label{vacuum-density}
\rho_{\mathrm{vac}} = \frac{1}{4 g_{YM}^2} B_{ab}^2.
\end{equation}
Thus the flat space $\mathbb{R}^{2n}$ is {\it emergent} from the uniform
vacuum condensate. This implies a remarkable picture [7, 8] that the flat space is not an empty space
unlike the general relativity but emergent from a coherent vacuum condensate corresponding to a cosmological
constant in general relativity. We will see soon that it is an inevitable consequence of emergent spacetime
and the emergent spacetime picture will completely change the situation
regarding to the cosmological constant problem.\footnote{The action density \eq{vacuum-density} in $2n$-dimensions
corresponds to the energy density for static solutions in $(2n+1)$-dimensions. Until we introduce the concept
of emergent time, we will discuss the cosmological constant problem in this context.}

In section 3, we have introduced a Lie algebra homomorphism $\rho: \mathcal{A}_\theta \to \mathfrak{D}$ where
$\mathfrak{D}$ denotes the vector space of inner derivations in Eq. \eq{der-d}.
In particular, the set of derivations defined by \eq{nc-vector} plays a fundamental role because
the matrix model depends only on the combination \eq{nc-fluct}.
Since every automorphism of the matrix algebra $\mathcal{A}_N$ is inner, a general element of
the automorphism in \eq{inner-auto} may be generated by
\begin{equation}\label{mat-auto}
    U_c = \exp \big( - ic^a \Phi_a \big) \in \mathcal{A}_N
\end{equation}
with $c^a \in \mathbb{R}$ or $\mathbb{C}$. Its infinitesimal generator leads to
the fundamental derivations in Eq. \eq{nc-vector}.
Substituting the fluctuations \eq{nc-fluct} into the Jacobi identity \eq{m-jacobi}
and the equations of motion \eq{m-eom} leads to the Bianchi identity and
the equations of motion of NC $U(1)$ gauge fields, respectively:
\begin{eqnarray} \label{nc-bianchi}
  && \widehat{D}_{[a} \widehat{F}_{bc]} \equiv \widehat{D}_{a} \widehat{F}_{bc}
+ \widehat{D}_{b} \widehat{F}_{ca} + \widehat{D}_{c} \widehat{F}_{ab} = 0, \\
\label{ncu1-eom}
  && \widehat{D}_{b} \widehat{F}_{ab} = 0,
\end{eqnarray}
where we used the relation
\begin{equation}\label{double-comm}
    [\Phi_a, [\Phi_b, \Phi_c]] = - \widehat{D}_a \widehat{F}_{bc}.
\end{equation}
Using the relation \eq{double-comm} and the algebra homomorphism \eq{lie-homo}, one can get the derivations
\begin{equation}\label{double-der}
    \mathrm{ad}_{\widehat{D}_a \widehat{F}_{bc}} = [\widehat{V}_a, [\widehat{V}_b, \widehat{V}_c]].
\end{equation}
Using this relation, the Bianchi identity \eq{nc-bianchi} and the equations of motion \eq{ncu1-eom}
for NC $U(1)$ gauge fields are mapped to algebraic (eventually geometric) equations of derivations
in $\mathfrak{D}$:
\begin{eqnarray} \label{bianchi-geom}
  && \widehat{D}_{[a} \widehat{F}_{bc]} = 0 \quad \Longrightarrow  \quad
[\widehat{V}_a, [\widehat{V}_b, \widehat{V}_c]] + [\widehat{V}_b, [\widehat{V}_c, \widehat{V}_a]]
+ [\widehat{V}_c, [\widehat{V}_a, \widehat{V}_b]] = 0, \\
\label{eom-geom}
  && \widehat{D}_{b} \widehat{F}_{ab} = 0 \quad \Longrightarrow \quad
[\widehat{V}_b, [\widehat{V}_a, \widehat{V}_b]] = 0.
\end{eqnarray}

The vector space $\mathfrak{D}$ of derivations is realized as differential operators acting on $C^\infty(N)$.
Hence it is convenient to use the $*$-product representation \eq{*-product} of
the NC algebra $\mathcal{A}_\theta$. Then the generalized vector fields $\mathrm{ad}_{\widehat{\phi}_a} = \widehat{V}_a$
in Eq. \eq{nc-vector} take the form \cite{q-emg,hsy-inflation}
\begin{equation}\label{polyvector}
  \widehat{V}_a = V_a^\mu (y) \frac{\partial}{\partial y^\mu} + \sum_{p=2}^\infty
  V^{\mu_1 \cdots \mu_p}_a (y) \frac{\partial}{\partial y^{\mu_1}} \cdots
  \frac{\partial}{\partial y^{\mu_p}} \in \mathfrak{D}.
\end{equation}
One can see that the generalized vector fields in $\mathfrak{D}$ generate an infinite tower of
the so-called polyvector fields. Note that the covariant momenta $\widehat{\phi}_a$ and
$\widehat{\phi}'_a = \widehat{\phi}_a + \widehat{D}_a \widehat{\lambda}$ for $\widehat{\lambda}
\in \mathcal{A}_\theta$ are in the same gauge equivalence class. Therefore the corresponding polyvector fields
in $\mathfrak{D}$ should be identified within the equivalence classes defined by
\begin{equation}\label{equiv-d}
\widehat{V}_a \sim \widehat{V}'_a = \widehat{V}_a + [\widehat{V}_a, \widehat{V}_{\widehat{\lambda}}]
\end{equation}
where $\widehat{V}_{\widehat{\lambda}} = \mathrm{ad}_{\widehat{\lambda}}$.
It is important to perceive that the realization of $\mathfrak{D}$ through the derivation algebra
in Eq. \eq{polyvector} is intrinsically local \cite{q-emg}. Therefore it is necessary to consider patching
or gluing together the local constructions using the gauge degrees of freedom \eq{equiv-d}
to form a set of global quantities. We will assume that local coordinate patches have been consistently
glued together to yield global polyvector fields.

In a large distance limit, i.e., $|\theta| \to 0$, we expect to recover a K\"ahler geometry from
a ``quantum geometry" according  to the duality picture \eq{q-diag}. In other words,
we need to show that a K\"ahler geometry is emergent from the commutative limit of
the zero-dimensional matrix model \eq{matrix-action} in the NC Coulomb branch.
Suppose that the resulting K\"ahler geometry is described by $(M, g)$ with the metric \eq{c-metric}.
In that limit, the polyvector fields in \eq{polyvector} reduce to
ordinary vector fields that will be identified with frame fields in $\Gamma(TM)$.
Let us denote the globally defined vector fields by
\begin{equation}\label{global-vec}
    \mathfrak{X}(M) = \left\{ V_a = V_a^\mu (x) \frac{\partial}{\partial x^\mu}| a, \mu = 1, \cdots, 2n \right\}.
\end{equation}
Define the structure equations of vector fields by
\begin{equation}\label{str-eq}
    [V_a, V_b] = - {g_{ab}}^c V_c.
\end{equation}
The orthonormal vielbeins on $TM$ are then defined by the relation $V_a = \lambda E_a \in \Gamma(TM)$ where
a positive function $\lambda$ is to be determined by a volume-preserving condition.
We fix the conformal factor $\lambda$ by imposing the condition that the vector fields $V_a$ preserve a volume form
\begin{equation}\label{volume-v}
    \nu = \lambda^2 v^1 \wedge \cdots \wedge v^{2n}
\end{equation}
where $v^a = v^a_\mu (x) dx^\mu \in \Gamma(T^* M)$ are covectors dual to $V_a$, i.e.,
$\langle v^a, V_b \rangle = \delta^a_b$.
This means that the vector fields $V_a$ obey the condition
\begin{equation}\label{vol-cond}
    \mathcal{L}_{V_a} \nu = \big ( \nabla \cdot V_a + (2-2n) V_a \ln \lambda \big) \nu = 0.
\end{equation}
The above condition \eq{vol-cond} can be written as
\begin{equation}\label{ivol-cond}
    \frac{1}{\sqrt{\det g_V}} \partial_\mu \big( V_a^\mu \sqrt{\det g_V} \big)
+ 2 V_a \ln \lambda = 0     \qquad \Leftrightarrow \qquad
{g_{ba}}^b = V_a \ln \lambda^2,
\end{equation}
where $\sqrt{\det g_V} = \det v^a_\mu$. If the vector fields $V_a$ are known, the conformal factor $\lambda^2$
can be determined by solving Eq. \eq{ivol-cond}. Then the Riemannian metric on an emergent K\"ahler manifold
is completely determined by
\begin{equation}\label{ek-metric}
    ds^2 = e^a \otimes e^a = \lambda^2 v^a \otimes v^a = \lambda^2 v^a_\mu v^a_\nu dx^\mu \otimes dx^\nu
\end{equation}
where $e^a = \lambda v^a \in \Gamma(T^* M)$ are the orthonormal coframes.
From the definition \eq{vol-cond}, one can see that the factor $\lambda$ can be determined only up to
a constant scaling, $\lambda \to \beta \lambda$, where $\beta \in \mathbb{R} > 0$.
The scaling freedom can be attributed to the scale symmetry of the dynamical variables,
$\Phi_a \to \beta \Phi_a$, in \eq{nc-fluct}. This scale symmetry will be fixed by relating
the dynamical scale of the vacuum condensate \eq{nc-coulomb} to a characteristic scale of quantum gravity.

Given a connection $\nabla$ on $TM$ for any Riemannian manifold $M$,
the torsion and the curvature are defined by \cite{besse}
\begin{eqnarray} \label{def-tor}
&& T(X, Y) = \nabla_X Y - \nabla_Y X - [X, Y], \\
\label{def-cur}
&& R(X, Y) Z= [\nabla_X, \nabla_Y]Z -\nabla_{[X,Y]} Z,
\end{eqnarray}
where $X, Y$ and $Z$ are vector fields on $M$.
They are multi-linear differential operators, i.e.,
\begin{eqnarray} \label{lin-tor}
&& T(f X, Y) = T(X, f Y) = f T(X, Y), \\
\label{lin-cur}
&& R(f X, Y) Z= R(X, f Y) Z = R(X, Y) f Z = f R(X, Y) Z,
\end{eqnarray}
where $f \in C^\infty (M)$. The connection is torsion-free if $T(X, Y) = 0$ for all $X, Y \in \mathfrak{X}(M)$,
that is $\nabla_X Y - \nabla_Y X = [X, Y]$.
Then one can easily show that
\begin{equation}\label{r-jacobi}
 R(X, Y) Z + R(Y, Z) X + R(Z, X) Y = [X, [Y, Z]] + [Y, [Z, X]] + [Z, [X, Y]]
\end{equation}
for any $X, Y, Z \in \mathfrak{X}(M)$. In our case, Eqs. \eq{lin-tor} and \eq{lin-cur} imply that
$T(V_a, V_b) = \lambda^2 T(E_a, E_b)$ and $R(V_a, V_b) V_c = \lambda^3 R(E_a, E_b) E_c$.
In particular, after imposing the torsion free condition, $T(E_a, E_b)=0$,
Eq. \eq{r-jacobi} gives us the relation \cite{hsy-jhep09}
\begin{equation}\label{1st-jacobi}
R(E_{[a}, E_b) E_{c]} = \lambda^{-3} R(V_{[a}, V_b) V_{c]} = \lambda^{-3} [V_{[a}, [V_b,V_{c]}]]
\end{equation}
where the symbol $[\cdots]$ denotes the cyclic permutation of indices inside of it.

Since the set of vector fields in \eq{global-vec} arises at a commutative limit,
$|\theta| \to 0$, of the generalized vector fields \eq{polyvector}, Eqs. \eq{bianchi-geom}
and \eq{eom-geom} give rise to the following correspondence, respectively,
\begin{eqnarray} \label{cor-bianchi}
  &&  \widehat{D}_{[a} \widehat{F}_{bc]} = 0 \quad \xRightarrow{|\theta| \to 0} \quad
  [V_{[a}, [V_b,V_{c]}]] = 0, \\
\label{cor-eom}
  &&  \widehat{D}_b \widehat{F}_{ab} = 0 \quad \xRightarrow{|\theta| \to 0} \quad
  [V_b, [V_a, V_b] ] = 0.
\end{eqnarray}
Consequently, Eq. \eq{1st-jacobi} shows \cite{hsy-jhep09,hsy-review} that the Bianchi identity \eq{cor-bianchi}
for NC $U(1)$ gauge fields in the commutative limit is equivalent to the first Bianchi identity
for the Riemann curvature tensors, i.e.,
\begin{equation}\label{bianchi-ncgr}
  \widehat{D}_{[a} \widehat{F}_{bc]} = 0 \quad \xRightarrow{|\theta| \to 0} \quad  R(E_{[a}, E_b) E_{c]} =0.
\end{equation}
The underlying argument leading to the result \eq{bianchi-ncgr} implies that the correspondence will be true
for general cases beyond a K\"ahler geometry as far as a fundamental algebra is associative, e.g. \eq{m-jacobi}.
The mission for the equations of motion \eq{cor-eom} is more involved. But, from the experience
on the Bianchi identity \eq{bianchi-ncgr}, we basically expect that it will be reduced
to the Einstein equations
\begin{equation}\label{eom-ncgr}
  \widehat{D}_{b} \widehat{F}_{ab} = 0 \quad \xRightarrow{|\theta| \to 0} \quad
  R_{ab} = 8 \pi G \Big( T_{ab} - \frac{1}{2} \delta_{ab} T \Big).
\end{equation}
Now we will show that the correspondence for the Einstein equations \eq{eom-ncgr} is true at least
for K\"ahler manifolds.

In the commutative limit, the right-hand side of Eq. \eq{cor-eom} can be written as
\begin{equation}\label{cl-eom}
    V_c g_{cab} - g_{cad} g_{cdb} = 0
\end{equation}
using the structure equation \eq{str-eq}. Contracting free indices in \eq{cl-eom} and
using Eq. \eq{ivol-cond} leads to the relation
\begin{equation}\label{eom-scalar}
 V_c^2 \ln \lambda = - \frac{1}{2} g_{cab} g_{cba}.
\end{equation}
After a little algebra, one can show \cite{hsy-jhep09,hsy-review} that
\begin{eqnarray} \label{eom-einstein}
  && - \frac{1}{2 \lambda^2}\big( V_c g_{cab} + V_c g_{cba} - g_{cad} g_{cdb} - g_{cbd} g_{cda} \big) \\
 &=& R_{ab} - \frac{1}{\lambda^2} \Big( \frac{4-d}{2} (V_a V_b + V_b V_a) \ln \lambda - \delta_{ab} V_c^2 \ln \lambda
  + (d-2) V_a \ln \lambda V_b \ln \lambda -  (d-4) \delta_{ab} \big( V_c \ln \lambda \big)^2  \xx
  && + \frac{d-4}{2} \big( g_{acb} + g_{bca} \big) V_c \ln \lambda
- \frac{1}{2} \big( g_{cad} g_{cdb} + g_{cda} g_{cbd} + g_{acd} g_{bdc} + g_{acd} g_{bcd} \big)
+ \frac{1}{4} g_{cda} g_{cdb} \Big), \nonumber
\end{eqnarray}
where $d=2n$. Since the first line in Eq. \eq{eom-einstein} has to vanish according to the equations
of motion \eq{cl-eom}, we get the Einstein equations as the form \eq{eom-ncgr}.
The remaining problem is to identify the energy-momentum tensor $T_{ab}$ determined by NC $U(1)$ gauge fields.
One may notice that $T_{ab}$ can be written as the products of the structure function $g_{abc}$ after using
Eqs. \eq{ivol-cond} and \eq{eom-scalar} except the term $(V_a V_b + V_b V_a) \ln \lambda$.

In four dimensions $(d=4)$, several terms in $T_{ab}$ vanish and the Einstein equations are given by
\begin{equation}\label{4d-einstein-eq}
R_{ab} = - \frac{1}{2 \lambda^2} \Big( g_{cad} g_{cdb} + g_{cda} g_{cbd} + g_{acd} g_{bdc} + g_{acd} g_{bcd}
- \frac{1}{2} g_{cda} g_{cdb} - g_{cac} g_{dbd} - \delta_{ab} g_{cde} g_{ced} \Big),
\end{equation}
where Eqs. \eq{ivol-cond} and \eq{eom-scalar} were used.
In order to understand the energy-momentum tensor, it is convenient to take the canonical decomposition
of the structure equation \eq{str-eq} as
\begin{equation}\label{dec-ste}
  g_{abc} = g^{(+)i}_c \eta^i_{ab} + g^{(-)i}_c \overline{\eta}^i_{ab}.
\end{equation}
Then the energy-momentum tensor $T_{ab}$ is given by a remarkably simple but cryptic result \cite{hsy-review,hsy-jhep09}
\begin{equation}\label{4d-ricci-emt}
    R_{ab} = - \frac{1}{\lambda^2} \Big( g^{(+)i}_d g^{(-)j}_d (\eta^i_{ac} \overline{\eta}^j_{bc}
    + \eta^i_{bc} \overline{\eta}^j_{ac}) - g^{(+)i}_c g^{(-)j}_d (\eta^i_{ac} \overline{\eta}^j_{bd}
    + \eta^i_{bc} \overline{\eta}^j_{ad}) \Big).
\end{equation}
Note that the first combination on the right-hand side of Eq. \eq{4d-ricci-emt} must be traceless
because of $\eta^i_{ab} \overline{\eta}^j_{ab} = 0$ while the second is not. Therefore, one can see that
the energy-momentum tensor induced by NC $U(1)$ gauge fields surprisingly gives rise to a nontrivial
Ricci scalar given by
\begin{equation}\label{4d-ricci-scalar}
    R= \frac{2}{\lambda^2} g^{(+)i}_b g^{(-)j}_c \eta^i_{ab} \overline{\eta}^j_{ac}
\end{equation}
although we have started with a pure NC $U(1)$ gauge theory without any other fields.
On the one hand, the first combination that gives rise to a traceless energy-momentum tensor
can be written as the form
\begin{equation}\label{maxwell-emt}
    8\pi G T_{ab}^{(M)} \equiv - \frac{1}{\lambda^2} \Big( g_{acd} g_{bcd}
- \frac{1}{4} \delta_{ab} g_{cde} g_{cde} \Big).
\end{equation}
On the other hand, the second combination that gives rise to the nontrivial Ricci scalar \eq{4d-ricci-scalar}
takes the form
\begin{equation}\label{liouville-emt}
    8\pi G T_{ab}^{(L)} \equiv \frac{1}{2 \lambda^2} \Big( \rho_a \rho_b - \Psi_a \Psi_b -
\frac{1}{2} \delta_{ab} (\rho^2_c  - \Psi^2_c ) \Big)
\end{equation}
in terms of the variables defined by
\begin{equation}\label{two-var}
    \begin{array}{l}
      \rho_a \equiv g_{bab} =  - \big( g^{(+)i}_b \eta^i_{ab} + g^{(-)i}_b \overline{\eta}^i_{ab} \big), \\
     \Psi_a \equiv - \frac{1}{2} \varepsilon^{abcd} g_{bcd} =  - \big( g^{(+)i}_b \eta^i_{ab}
- g^{(-)i}_b \overline{\eta}^i_{ab} \big).
    \end{array}
\end{equation}
The Ricci scalar \eq{4d-ricci-scalar} is then given by
\begin{equation}\label{4d-rscalar}
    R =  \frac{1}{2 \lambda^2} (\rho^2_a  - \Psi^2_a ).
\end{equation}

A close inspection of the energy-momentum tensors reveals many intriguing results.
First, note that the relation \eq{imp-exam} indicates the following map
$\rho: \mathcal{A}_\theta \to \mathfrak{D}$:
\begin{equation}\label{map-fcomm}
  \widehat{F}_{ab}  \qquad \xRightarrow{|\theta| \to 0} \qquad
  [V_a, V_b] = - {g_{ab}}^c V_c.
\end{equation}
Hence the commutative limit of NC $U(1)$ instantons obeying the self-duality
equations \eq{high-ncinst} in four dimensions is equivalently stated as
\begin{equation}\label{4d-nc-inst}
\left\{
  \begin{array}{ll}
    g^{(+)i}_a = 0, & \hbox{anti-self-dual;} \\
    g^{(-)i}_a = 0, & \hbox{self-dual}
  \end{array}
\right.
\end{equation}
for all $a, i$. In this case, $R_{ab} = 0$ according to Eq. \eq{4d-ricci-emt}.
Therefore NC $U(1)$ instantons generate no energy-momentum tensor as expected.
This means that the commutative limit of NC $U(1)$ instantons called symplectic $U(1)$ instantons
corresponds to Ricci-flat four-manifolds.
If we restrict NC $U(1)$ gauge fields to those arising from the quantization of
a local K\"ahler form in the sense of \eq{q-gauge2}, symplectic $U(1)$ instantons
must be gravitational instantons because a Ricci-flat, K\"ahler manifold is a gravitational instanton.
Therefore we confirm the picture \eq{cy-diag}.
One may use a natural norm in the vector space $\mathfrak{X}(M), \; (V_a \circ V_b) =
\lambda^2 (E_a \circ E_b) = \lambda^2 \delta_{ab}$, to define a corresponding operation
in the algebra $\mathcal{A}_\theta$. It was argued in \cite{hsy-jhep09,hsy-review} that
the energy-momentum tensor \eq{maxwell-emt} can be mapped to that of ordinary Maxwell theory
by simply interpreting the norm in the vector space $\mathfrak{X}(M)$ as
the product of Weyl symbols of operators in $\mathcal{A}_\theta$. If so, we see that the zero-dimensional
matrix model \eq{matrix-action} in the NC Coulomb branch is mapped in the commutative limit
to Einstein-Maxwell theory coupled to a \textit{mystical} energy-momentum tensor \eq{liouville-emt}.
One may decouple the energy-momentum tensor \eq{liouville-emt}
by considering field configurations satisfying the relation, $\rho_a = \pm \Psi_a$, which can be written as
\begin{equation}\label{scalar-inst}
\left\{
  \begin{array}{ll}
    \rho_a = + \Psi_a \;\; \Rightarrow \;\; \overline{\eta}^i_{ab}  g^{(-)i}_b  = 0, \\
   \rho_a = - \Psi_a \;\; \Rightarrow \;\; \eta^i_{ab}  g^{(+)i}_b  = 0.
  \end{array}
\right.
\end{equation}
In this case, the energy-momentum tensor \eq{liouville-emt} as well as the Ricci scalar \eq{4d-rscalar}
identically vanishes but the Maxwell energy-momentum tensor \eq{maxwell-emt} is nontrivial.
Thus the solution satisfying Eq. \eq{scalar-inst} describes a scalar-flat K\"ahler manifold
in Einstein-Maxwell theory whose general solutions
with a $U(1)$ isometry have been constructed in \cite{lebrun} (see also \cite{bubble-geom}).
However, it seems to be difficult to write down Eq. \eq{scalar-inst} as a local form in terms of symplectic
$U(1)$ gauge fields \cite{bottom-up}.

Now we are ready to discuss how the cosmological constant problem can be resolved in emergent gravity.
First of all, it should be instructive to trace the metric \eq{ek-metric} out to see where
the flat space $\mathbb{R}^{2n}$ comes from.
Definitely the flat space corresponds to the vector field $V_a = \delta^\mu_a \frac{\partial}{\partial x^\mu}$
for which $\lambda^2 = 1$. It is easy to see that the flat space arises from the vacuum
gauge fields $\langle \phi_a \rangle_{\mathrm{vac}} = p_a$,
defining the NC Coulomb branch in Eq. \eq{nc-coulomb}.
In this case, the automorphism \eq{g-auto} precisely corresponds to the Poincar\'e symmetry of
$\mathbb{R}^{2n}$ because the Coulomb branch vacua connected by \eq{g-auto} equally generate
the flat space. It means that the vacuum algebra generating the flat space $\mathbb{R}^{2n}$
is not unique but degenerate up to the automorphism \eq{g-auto}.
Therefore the NC Coulomb branch \eq{nc-coulomb} does not break the Poincar\'e symmetry.
Rather it is emergent from the symmetry \eq{g-auto} of the underlying matrix model.
It should be appreciated by recalling that the K\"ahler form of $\mathbb{R}^{2n}$ is given by
the two-form $B$ in Eq. \eq{kahler-gauge}. Thus the gauge theory formulation of K\"ahler geometry
clearly illuminates why it is necessary to have a vacuum condensate in order to generate even a flat space.
As we pointed out in Eq. \eq{vacuum-density}, the NC Coulomb branch \eq{nc-coulomb} actually generates
a huge vacuum energy of order $\rho_{\mathrm{vac}} \sim M_P^4$ since the Planck energy $M_P$ is
the natural dynamical scale for the generation of space(time). However the vacuum energy is simply
used to generate a flat space from \textit{nothing} and does not contribute to
the energy-momentum tensor as one can see from Eq. \eq{eom-einstein}. This property is a general
feature in emergent gravity.
As we pointed out in section 3, the derivation $\mathfrak{D}$ is inert for elements of the center of
the algebra $\mathcal{A}_\theta$ denoted by $\mathcal{Z} (\mathcal{A}_\theta)$, i.e., for an observable
$\mathcal{O} \in \mathcal{A}_\theta$,
\begin{equation}\label{inert}
  \mathrm{ad}_{\mathcal{O}} = \mathrm{ad}_{\mathcal{O} + \mathcal{I}}
\end{equation}
if $\mathcal{I} \in \mathcal{Z} (\mathcal{A}_\theta)$.
And the vacuum energy belongs to the center $\mathcal{Z} (\mathcal{A}_\theta)$.
Thus the emergent gravity is completely immune from the vacuum energy. In other words, the vacuum
energy \textit{does not gravitate} unlike to Einstein gravity. This is an underlying logic why the
emergent gravity can resolve the cosmological constant problem \cite{hsy-jhep09}.

Since the energy-momentum tensor \eq{liouville-emt} contributes a nontrivial Ricci scalar
to a K\"ahler manifold, it will be interesting to understand under what circumstances
this plays an important role. It is naturally expected that its scalar component
dominates at large distance scales. In Lorentzian spacetime, this scalar mode will cause an expansion
or contraction of spacetime while quadruple (symmetric and traceless) modes generate a shear distortion.
Therefore it is more instructive to address the issue in the Lorentzian spacetime.
In order to get a corresponding result in (3+1)-dimensional Lorentzian spacetime,
let us take the analytic continuation defined by $x^4 = ix^0$.
Under this Wick rotation, $g_{\mu\nu} \to g_{\mu\nu}, \; \rho_\mu \to \rho_\mu$ and $\Psi_\mu \to i \Psi_\mu$,
the Liouville energy-momentum tensor and the Ricci scalar in the Lorentzian signature are given by
\begin{eqnarray}\label{lorentz-emt}
&& T_{\mu\nu}^{(L)} = \frac{1}{16 \pi G \lambda^2} \Big(\rho_{\mu} \rho_{\nu} + \Psi_\mu \Psi_\nu
 - \frac{1}{2} g_{\mu\nu} (\rho^2_{\lambda} + \Psi^2_\lambda) \Big), \\
\label{lorentz-r}
&& R = \frac{1}{2 \lambda^2}  g^{\mu\nu} (\rho_{\mu} \rho_{\nu} + \Psi_\mu \Psi_\nu).
\end{eqnarray}
Now $\rho_\mu$ and $\Psi_\mu$ are four-vectors and random fluctuations in nature.
The Lorentzian four-vectors have their own causal structure unlike the Riemannian case.
Thus the Lorentzian spacetime is in sharp contrast to the Riemannian space
where every vectors are positive-definite.
They are classified into three classes: space-like, time-like, and null vectors for which
the Ricci scalar \eq{lorentz-r} becomes positive, negative and zero, respectively.
Hence their causal structure results in the different nature of gravitational interactions.
Indeed it can be shown \cite{dmde} that space-like fluctuations give rise
to the repulsive gravitational force while time-like fluctuations generate the attractive gravitational force.
When considering the fact that the fluctuations are random in nature and we are living
in the (3+1)-dimensional spacetime, the ratio of the repulsive and attractive components
will end in $\frac{3}{4}: \frac{1}{4}=75:25$
and this ratio curiously coincides with the dark composition of our current Universe.
It was argued in \cite{dmde} that the emergent gravity can explain the dark sector of our Universe
more precisely if one includes ordinary matters which act as the attractive force.

Thus we see that dark energy and dark matter would not be understood by simply modifying
the general relativity and quantum field theories. Another novel paradigm, a.k.a. quantum gravity,
is necessary to understand the nature of dark energy and dark matter.
After some contemplation, we are driven to a conclusion that the background-independent formulation of
quantum gravity through the concept of emergent spacetime is a core reason why the matrix model
provides a novel perspective to resolve the notorious problems in theoretical physics such as
the cosmological constant problem, dark energy, and dark matter.
Since the matrix model \eq{matrix-action} is a zero-dimensional theory,
any concept of space (and time) is not assumed in advance.
Only matrices (as objects) in $\mathcal{A}_N$ and their relationships (as morphisms) are at the first onset.
The coherent vacuum \eq{nc-coulomb} obeying the Heisenberg algebra is a crucial ingredient
for a macroscopically extended spacetime to be emergent from nothing.
(We will use spacetime as an abuse of the terminology since we will shortly discuss the concept
of emergent time.)
The spacetime vacuum in the NC Coulomb branch generates a constant vacuum energy \eq{vacuum-density}
whose characteristic dynamical scale is set by the Planck energy $M_P \sim 10^{18}$ GeV.
Therefore the vacuum spacetime (as a stage for fluctuations over there) behaves like a metrical elasticity
which opposes the curving of space. On the one hand, the gravitational force (as fluctuations
over the vacuum spacetime) will be extremely weak because the tension of spacetime is extremely large,
typically, of the Planck energy and so the space strongly withstands the curving.
Consequently, the dynamical origin of flat spacetime explains the metrical elasticity opposing
the curving of space and the stunning weakness of gravitational force \cite{hsy-jpcs12}.
Furthermore, as we argued around Eq. \eq{inert}, the emergent spacetime implies that
the global Lorentz symmetry, being an isometry of flat spacetime,
should be a perfect symmetry up to the Planck scale because the flat spacetime was originated
from the condensation of the maximum energy in Nature.
On the other hand, the vacuum algebra in the NC Coulomb branch \eq{nc-coulomb} is mathematically
the same as the Heisenberg algebra \eq{nc-phase} in quantum mechanics.
So there exists the spacetime uncertainty relation \eq{uncertain-nc} like as the Heisenberg's
uncertainty relation in quantum mechanics.
This implies that fluctuations of large $N$ matrices or
NC $U(1)$ gauge fields are not necessarily localized and some fluctuations can be extended
to macroscopic scales. This phenomenon is known as the UV/IR mixing or holography principle.
In other words, the NC Coulomb branch \eq{nc-coulomb}
satisfying the Heisenberg algebra necessarily gives rise to the UV/IR mixing
as a result of the spacetime uncertainty relation \eq{uncertain-nc}
and UV fluctuations at microscopic levels are necessarily paired with
IR fluctuations at macroscopic levels.
It was argued in \cite{dmde} that the UV/IR mixing is one of cruxes to understand the nature
of dark energy and dark matter from the emergent spacetime picture.
These macroscopic manifestations of quantum gravity effects would be the cornerstone
for the experimental verification of quantum gravity.

Let us also discuss how the zero-dimensional matrix model \eq{matrix-action} describes
six-dimensional K\"ahler manifolds in classical limit.
One can read off the Einstein equations in six dimensions $(d=6)$ from Eq. \eq{eom-einstein},
which are given by
\begin{eqnarray} \label{6d-einstein-eq}
 R_{ab} &=& \frac{1}{\lambda^2} \Big( - (V_a V_b + V_b V_a) \ln \lambda
  + 4 V_a \ln \lambda V_b \ln \lambda + \big( g_{acb} + g_{bca} \big) V_c \ln \lambda
- \delta_{ab} V_c^2 \ln \lambda \xx
&& - 2 \delta_{ab} \big( V_c \ln \lambda \big)^2
- \frac{1}{2} \big( g_{cad} g_{cdb} + g_{cda} g_{cbd} + g_{acd} g_{bdc} + g_{acd} g_{bcd} \big)
+ \frac{1}{4} g_{cda} g_{cdb} \Big).
\end{eqnarray}
Then one can read off the energy-momentum tensor and the Ricci scalar from Eq. \eq{6d-einstein-eq}:
\begin{eqnarray} \label{6d-em}
8 \pi G T_{ab} &=& \frac{1}{\lambda^2} \Big( - (V_a V_b + V_b V_a) \ln \lambda
  + 4 V_a \ln \lambda V_b \ln \lambda + \big( g_{acb} + g_{bca} \big) V_c \ln \lambda \xx
&& + \frac{1}{4} \delta_{ab} V_c^2 \ln \lambda - \delta_{ab} \big( V_c \ln \lambda \big)^2
- \frac{1}{2} \big( g_{cad} g_{cdb} + g_{cda} g_{cbd} + g_{acd} g_{bdc} + g_{acd} g_{bcd} \big) \xx
&& + \frac{1}{4} g_{cda} g_{cdb} + \frac{1}{16} \delta_{ab} g_{cde} g_{cde} \Big), \\
\label{6d-r-scalar}
R &=& - \frac{1}{\lambda^2} \Big( 5 V^2_a \ln \lambda + 4 \big( V_a \ln \lambda \big)^2
+ \frac{1}{4} g_{abc} g_{abc} \Big).
\end{eqnarray}

In six dimensions, the energy-momentum tensor \eq{6d-em} is poorly understood,
so let us focus on the self-dual case \eq{high-ncinst}. In the classical limit,
the operator algebra describes the so-called symplectic Hermitian $U(1)$ instantons
which are mapped using the correspondence \eq{map-fcomm} to the structure equations
\begin{equation}\label{shu1-inst}
    {g_{ab}}^e = \frac{1}{2} T_{abcd} {g_{cd}}^e.
\end{equation}
After some algebra, the above equations can be translated into the relationship of
$SO(6) \cong SU(4)$ spin connections \cite{hsy-cy}:
\begin{equation}\label{spin-inst}
    {\omega^e}_{ab} = \frac{1}{2} T_{abcd} {\omega^e}_{cd}.
\end{equation}
This means that the spin connections derived from symplectic Hermitian $U(1)$ instantons must
take values in $su(3)$ Lie algebra. This result leads to the conclusion that
symplectic Hermitian $U(1)$ instantons correspond
to Calabi-Yau manifolds because the latter is the K\"ahler manifolds with $SU(3)$ holonomy.
However it is nontrivial to show the Ricci-flatness of symplectic Hermitian $U(1)$ instantons directly
from the expression \eq{6d-einstein-eq}. Actually it is necessary to take several technical steps
to show $R_{ab}=0$ directly using Eq. \eq{shu1-inst}. In particular, it is required to use
the gauge degree of freedom in \eq{equiv-d}. In the four-dimensional case, 't Hooft symbols provide
a useful decomposition \eq{dec-ste} of antisymmetric tangent indices
into $SU(2)_L$ and $SU(2)_R$ parts \cite{riem-ym}.
Similarly, the six-dimensional 't Hooft symbols found in \cite{ahep-yy} will provide
a convenient decomposition of antisymmetric $SO(6) \cong SU(4)$ indices into $SU(3), \; \mathbb{C}\mathbb{P}^3 =
SU(4)/U(3)$ and $U(1)$ parts. Using this decomposition, one may get some useful information about
the energy-momentum tensor \eq{6d-em}. This result will be reported elsewhere.

\section{Discussion}

So far we have considered only symplectic manifolds associated with K\"ahler manifolds in Euclidean space
and have derived a zero-dimensional IKKT-type matrix model \eq{map-fcomm} as a Hilbert space representation
of quantized K\"ahler manifolds. There are several directions for the generalizations beyond K\"ahler manifolds.
First of all, it is desirable to understand Lorentzian manifolds as a dynamical spacetime in general relativity.
For this purpose, we need to introduce the concept of (emergent) time which is a notorious issue
in quantum gravity. We will be simple-minded to avoid some subtle issues about the emergent time.
We will argue that the quantization of Lorentzian manifolds is described by the BFSS-type matrix model.
As another generalizations, on the one hand, one may relax the symplectic condition on the existence of
a nondegenerate, closed two-form $B$ on $N$. Since a symplectic manifold is a Poisson manifold $(N, \theta)$ with a
nondegenerate bi-vector field $\theta \in \Gamma(\Lambda^2 TN)$, one may consider a
Poisson manifold $(N, \theta)$ with a general Poisson bi-vector field $\theta \in \Gamma(\Lambda^2 TN)$ which
could be degenerate on a subspace in $N$. For example, the Poisson structure
$\theta^{ab} (x) = l_s {f^{ab}}_c x^c \; (a,b,c =1, \cdots,r)$ becomes degenerate at the origin.
In this case, a vacuum algebra is given by a Lie algebra
\begin{equation}\label{lie-vacuum}
    [x^a, x^b] = i l_s {f^{ab}}_c x^c.
\end{equation}
It can be shown that the Lie algebra vacuum \eq{lie-vacuum} arises in a massive matrix model.
On the other hand, one may relax the closedness condition on the existence of
the symplectic two-form $B$ on $N$.
An important example is locally conformal symplectic (LCS) manifolds \cite{lcs,vaisman}.
An LCS manifold is a triple $(N, B, b)$ where $b$ is a closed one-form and
$B$ is a nondegenerate (but not closed) two-form satisfying
\begin{equation}\label{lcs}
    dB = b \wedge B.
\end{equation}
It was shown in \cite{hsy-2016} that cosmic inflation is realized as an LCS manifold and it arises
as a time-dependent vacuum in the BFSS-type matrix model.

The theory of relativity dictates that space and time must be coalesced into the form of
Minkowski spacetime in a locally inertial frame.
Hence, if general relativity is realized from a NC algebra, it is necessary to put space
and time on an equal footing in the NC algebra. If space is emergent, so should time.
Thus, an important problem is how to realize the emergence of time from the algebraic approach.
Quantum mechanics offers us a valuable lesson that the definition of (particle) time is strictly
connected with the problem of dynamics called an evolution of a system.
To appreciate such an evolution of a system, first observe that the matrix model
\eq{matrix-action} allows infinitely many (probably uncountably many) solutions since arbitrary deformations
of the Heisenberg vacuum in Eq. \eq{nc-fluct} are again solutions as far as they satisfy \eq{ncu1-eom}.
Therefore one may parameterize the deformations as a one-parameter family according to
the matrix automorphism \eq{inner-auto}:
\begin{equation}\label{t-family}
    A(t) \equiv \Psi (A) = U_t A U_t^{-1}
\end{equation}
where $t$ is a real affine parameter. If $U_t = e^{i H t}$ and $H \in \mathcal{A}_N$,
the infinitesimal change for $0 < t \ll 1$ is given by
\begin{equation}\label{hamilton-inner}
\frac{d A(t)}{dt} = i [H, A].
\end{equation}
This implies that the $t$-evolution \eq{t-family} can be understood as an another inner automorphism
of the matrix algebra $\mathcal{A}_N$.
Thus the inner automorphism generated by $U_c$ in Eq. \eq{mat-auto} may be generalized as
\begin{equation}\label{gen-inner}
    U_C = \exp \big( i c^0 H  - i c^a \Phi_a \big) \in \mathcal{A}_N.
\end{equation}
As we have learned from quantum mechanics, the time-evolution of a dynamical system is more general
when the system is open and interacting with environments.
In this case, the time-evolution is generated by outer as well as inner automorphisms.
Thus the $t$-evolution \eq{t-family} must be replaced by
\begin{equation}\label{hamilton-inner}
\frac{d A(t)}{dt} = i [H, A] + \frac{\partial A}{\partial t} =
\Big( \frac{\partial}{\partial t} + \mathrm{ad}_{A_0} \Big) A
\end{equation}
where $A_0 \equiv - H \in \mathcal{A}_N$.
One may rewrite Eq. \eq{hamilton-inner} as
\begin{equation}\label{hamilton-ad}
\frac{d A(t)}{dt} = \mathrm{ad}_{\Phi_0} A = - i [\Phi_0, A] \equiv D_0 A
\end{equation}
by regarding $\Phi_0 \equiv  i \frac{\partial}{\partial t} + A_0$ as
a differential operator satisfying the Leibniz rule.
Let us represent an element in the general automorphism of $\mathcal{A}_N$ including
the outer automorphism in Eq. \eq{hamilton-inner} by
\begin{equation}\label{gen-inner}
    U^t_C = \exp \big(- i C^A \Phi_A \big) \in \mathrm{Aut}(\mathcal{A}^t_N),
\end{equation}
where $\Phi_A = (\Phi_0, \Phi_a)$ and $\mathcal{A}^t_N$ denotes the set of
time-dependent $N \times N$ matrices.

It may be obvious how to generalize the zero-dimensional matrix model \eq{matrix-action}
to time-dependent matrices in $\mathcal{A}^t_N$. Since we want to treat the generators
$\Phi_A = (\Phi_0, \Phi_a)$ in $\mathrm{Aut}(\mathcal{A}^t_N)$ on an equal footing,
we take the time-dependent matrix action for $\mathcal{A}^t_N$ as
\begin{eqnarray}\label{bfss-action}
    S &=& \frac{1}{4 g_1^2} \int dt \mathrm{Tr}_N \eta^{AC} \eta^{BD}
[\Phi_A, \Phi_B] [\Phi_C, \Phi_D] \xx
&=& \frac{1}{g_1^2} \int dt  \mathrm{Tr}_N \Big( \frac{1}{2} (D_0 \Phi_a)^2
+ \frac{1}{4} [\Phi_a, \Phi_b]^2 \Big)
\end{eqnarray}
where $g_1$ is a corresponding coupling constant and $\eta^{AB} = \mathrm{diag}
(-1, 1, \cdots, 1)$.\footnote{\label{lorentz}The reason why time direction appears
with an opposite norm to spatial ones is to keep both the kinetic term in the action and
the Hamiltonian $ H =  \frac{1}{2} (D_0 \Phi_a)^2
- \frac{1}{4} [\Phi_a, \Phi_b]^2$ positive definite.
Moreover it seems to be natural if time evolution is still well-defined under
the global automorphism $\Phi_A \to {L_A}^B \Phi_B + C_A$. Contrary to the Euclidean case \eq{g-auto},
it is not possible to completely interchange spatial and time directions using the automorphism
with the Lorentzian signature.}
This kind of one-dimensional matrix model has been suggested as a nonperturbative formulation of M-theory
in an infinite-momentum or light-cone frame \cite{bfss}. The equations of motion are given by
\begin{equation}\label{eom-bfss}
 - D^2_0 \Phi_a + [\Phi_b, [\Phi_a, \Phi_b]] = 0,
\end{equation}
which must be supplemented with the Gauss constraint
\begin{equation}\label{gauss}
 [ \Phi_a, D_0 \Phi_a] = 0.
\end{equation}

In the large $N \to \infty$ limit, the NC Coulomb branch is still a consistent vacuum satisfying
\eq{eom-bfss} and \eq{gauss}. Plugging the fluctuations \eq{nc-fluct} into the action \eq{bfss-action}
leads to the action of $(2n+1)$-dimensional NC $U(1)$ gauge fields given by
\begin{equation}\label{lm-action}
    S = - \frac{1}{4 \mathfrak{g}^2_{YM}} \int d^{2n+1} X \widehat{F}_{AB} \widehat{F}^{AB},
\end{equation}
where $X^A = (t, y^a), \; A=0,1, \cdots, 2n$ and $\widehat{F}_{AB} = \partial_A \widehat{A}_{B}
- \partial_B \widehat{A}_{A} - i [\widehat{A}_{A}, \widehat{A}_{B}]$ is the field strength
of NC $U(1)$ gauge fields $\widehat{A}_A = (\widehat{A}_0, \widehat{A}_a)$.
Denote the underlying NC algebra by $\mathcal{A}_\theta^t$.
One can apply the Lie algebra homomorphism $\rho_t: \mathcal{A}_\theta^t \to \mathfrak{D}_t$
to time-dependent gauge fields in $\mathcal{A}_\theta^t$ to derive an emergent Lorentzian spacetime.
It can be shown \cite{hsy-inflation} that the Lorentzian spacetime emergent from $\mathcal{A}_N^t$ or
$\mathcal{A}_\theta^t$ is asymptotically (locally) flat, i.e., locally approaches
to the vacuum spacetime $\mathbb{R}^{2n,1}$ at asymptotic infinities.
It turns out that the global automorphism in footnote \ref{lorentz} corresponds
to the Poincar\'e symmetry of Minkowski spacetime $\mathbb{R}^{2n,1}$.

Unfortunately it is difficult to realize, for instance, $(3+1)$-dimensional spacetime from
the matrix model \eq{bfss-action} by considering the NC Coulomb branch satisfying
the Heisenberg algebra \eq{nc-coulomb}.
In section 4, we have obtained the $(3+1)$-dimensional Lorentzian spacetime
by applying the analytic continuation $x^4 = i t$ to a Riemannian manifold derived
from the zero-dimensional matrix model.
We do not think that the analytic continuation to the Lorentzian signature is a big deal
so that it invalidates the argument about dark energy and dark matter.
Nevertheless, it should be desirable to find a matrix model description of our Universe
directly starting with the matrix model \eq{bfss-action}. Or one may make a detour by considering
a contact structure instead of symplectic structure.
For example, one may introduce the analytic continuation $t = - i y^{2n+1}$ and a contact structure
along $y^{2n+1}$-direction to realize $(2n+1)$-dimensional Riemannian manifolds
from the matrix model \eq{bfss-action} in the NC Coulomb branch satisfying the Heisenberg algebra \eq{nc-coulomb}.
Similarly, the contact structure can be used to formulate even- and odd-dimensional
Lorentzian manifolds from an underlying matrix model.
In particular, even-dimensional Lorentzian manifolds like our Universe may be derived
from a $(1+1)$-dimensional matrix model by using two (spatial and temporal) contact structures \cite{hsy-inflation}.

It is easy to show that the Lie algebra \eq{lie-vacuum} cannot be realized as a vacuum solution
in the matrix model \eq{matrix-action} or \eq{bfss-action} because
$[x^b, [x^a, x^b]] = - l_s^2 x^a$ for a simple Lie algebra.\footnote{One may notice that there is
a time-dependent vacuum given by $\Phi_a = \frac{m}{l_s} x^a \sin (mt + \alpha)$, which obeys
both \eq{eom-bfss} and \eq{gauss}. We restrict the vacuum to a stable, time-independent solution.}
It suggests that the matrix model has to contain
a mass term $\frac{1}{2} m^2 \Phi_a \Phi_a$ to realize such a Lie algebra vacuum.
In this case the equations of motion are replaced by
\begin{equation}\label{mass-eom}
    [\Phi_b, [\Phi_a, \Phi_b]] + m^2 \Phi_a = 0.
\end{equation}
One can check that the vacuum
 \begin{equation}\label{lie-massvac}
    \langle \Phi_a \rangle_{\mathrm{vac}} = \frac{m}{l_s} x^a
 \end{equation}
satisfies the above equations of motion. Since the vacuum moduli \eq{lie-massvac} satisfy a compact Lie algebra,
the vacuum manifold derived from them is in general a compact manifold.
For example, if they satisfy the $so(3)$ or $su(2)$ Lie algebra,
the corresponding vacuum manifold is a fuzzy sphere
for a large but finite $k$-dimensional representation \cite{fuzzys2,ourfuzzy}.
Thus the Riemannian manifold emergent from
vacuum matrices may have a non-vanishing Ricci scalar $R$ of order $\sim m^2$.
If the vacuum satisfies a semi-simple Lie algebra, the corresponding vacuum manifold will be given
by a direct product of compact manifolds for simple elements.

It is well-known \cite{georgi} that the Lie algebra generators for a simple Lie algebra can be
represented by quadratic forms in terms of creation/annihilation operators in
the Heisenberg algebra \eq{ho-algebra}. It is known as the Schwinger representation.
For instance, the Schwinger representation of Lie algebra generators for $G = SU(n)$ is given by
\begin{equation}\label{schwinger}
    x^a = a_i^\dag T^a_{ij} a_j, \qquad i, j= 1,\cdots,n, \quad a = 1, \cdots, r
\end{equation}
where $ r = \dim (G) = n^2 -1$ and $T^a$'s are constant $n \times n$ Hermitian matrices satisfying
the $su(n)$ Lie algebra, $[T^a, T^b] = i {f^{ab}}_c T^c$. Thus the Lie algebra generators \eq{schwinger}
are composite operators like as the angular momentum operators in quantum mechanics.
A notable point is that the angular momentum operators in quantum mechanics arise as symmetry operators
of hydrogen atom rather than fundamental dynamical variables.
Similarly the $su(n)$ generators in \eq{schwinger} may also arise as low-energy order parameters of
a topological object such as NC $U(1)$ instantons in \eq{cy-diag}. It was argued in \cite{q-emg} that
$SU(n)$ gauge fields for $n=2, 3$ in Standard Model may arise in this way if a six-dimensional internal space
is made of a Calabi-Yau $n$-fold or equivalently NC $U(1)$ instantons are formed along the six-dimensional
NC internal space according to the duality \eq{cy-diag}. Of course, this problem immediately accompanies
an important question: How to realize quarks and leptons in this context?
This problem eventually points to the issue of emergent quantum mechanics addressed in section 1.

As we pointed out in section 1, the NC space \eq{nc-space-int} is mathematically equivalent to
the NC phase space \eq{nc-phase}. Therefore the underlying math will be the same for both spaces.
The Liouville theorem in classical mechanics states that the volume of phase space occupied
by particles is invariant under time evolution for divergenceless Hamiltonian flows.
This is almost true for the vector field $X$ satisfying Eq. \eq{darboux-2form}, i.e.,
$\mathcal{L}_X B = F$. One can see that $X$ becomes a Hamiltonian vector field at spatial infinities
because $F=0$ there. This implies that the flow generated by $X$ leads to only local changes of
spatial volume while it preserves the volume element at asymptotic regions.
Hence the time evolution or flow generated by the vector field $X$ cannot produce an inflating spacetime
since cosmic inflation is the exponential expansion of space everywhere.
It is the reason why it is necessary to go beyond a symplectic manifold to generate the cosmic inflation
from the emergent gravity approach. Suppose that a manifold $(N, B)$ allows a vector field $X$ satisfying
\begin{equation}\label{conf-vec}
    \mathcal{L}_X B =  \kappa B
\end{equation}
where $\kappa \in \mathbb{R}^*$ is a nonzero real number. Such a vector field is called
a \textit{conformal} vector field and the LCS manifold $(N, B, b)$ defined by \eq{lcs} admits
the conformal vector fields \cite{vaisman}.
In this case, the flow $\phi_t$ generated by a conformal vector field
has the property $B_t \equiv \phi_t^* B = e^{\kappa t} B$. Then one can see that the cosmic inflation occurs
since the spatial volume is proportional to $B_t^n = e^{n \kappa t} B^n$.
It was shown in \cite{hsy-inflation} that the cosmic inflation is triggered
by the condensate of Planck energy into vacuum and
the inflationary universe arises as a time-dependent solution of the matrix model \eq{bfss-action}
without introducing any inflaton field as well as an \textit{ad hoc} inflation potential.
The emergent spacetime picture admits a background-independent formulation so that the
inflation is responsible for the dynamical emergence of spacetime described by
a conformal Hamiltonian system.

\section*{Acknowledgments}

This work was supported by the National Research Foundation of Korea (NRF) grant funded
by the Korea government (MOE) (No. NRF-2015R1D1A1A01059710).


\newpage

\begin{thebibliography}{55}


\bibitem{est-review} G. T. Horowitz and J. Polchinski, {\it Gauge/gravity duality}, [gr-qc/0602037];
N. Seiberg, {\it Emergent spacetime}, [hep-th/0601234]; J. Maldacena, {\it The Gauge/gravity duality},
[arXiv:1106.6073].



\bibitem{witten-se} E. Witten, {\it Symmetry and Emergence}, Nature Phys. {\bf 14} (2018) 116, [arXiv:1710.01791].



\bibitem{dirac} P. A. M. Dirac, {\it The principle of quantum mechanics}, 4th ed., Oxford Univ. Press (1958).



\bibitem{snyder-yang} H. S. Snyder, {\it Quantized Space-Time}, Phys. Rev. {\bf 71} (1947) 38;
C. N. Yang, {\it The Electromagnetic Field In Quantized Space-Time}, {\it ibid.} {\bf 72} (1947) 874;
S. Doplicher, K. Fredenhagen and J. E. Roberts,
{\it The quantum structure of spacetime at the Planck scale and quantum fields},
Commun. Math. Phys. {\bf 172} (1995) 187, [hep-th/0303037].



\bibitem{hsy-review} J. Lee and H. S. Yang, {\it Quantum gravity from noncommutative spacetime},
J. Korean Phys. Soc. {\bf 65} (2014) 1754, [arXiv:1004.0745].


\bibitem{q-emg} H. S. Yang, {\it Quantization of emergent gravity},
Int. J. Mod. Phys. A {\bf 30} (2015) 1550016 [arXiv:1312.0580].



\bibitem{hsy-2016} H. S. Yang, {\it Emergent spacetime for quantum gravity}, Int. J. Mod. Phys. D {\bf 26}
(2016) 1645010 [arXiv:1610.00011].



	
\bibitem{dmde} J. Lee and H. S. Yang, {\it Dark Energy and Dark Matter in Emergent Gravity},
[arXiv:1709.04914].



\bibitem{hsy-inflation} H. S. Yang, {\it Emergent Spacetime and Cosmic Inflation I \& II}, [arXiv:1503.00712].


\bibitem{hsy-essay} H. S. Yang, {\it Emergent Spacetime: Reality or Illusion?}, [arXiv:1504.00464].




\bibitem{uv-ir} S. Minwalla, M. Van Raamsdonk and N. Seiberg, {\it Noncommutative perturbative dynamics},
J. High Energy Phys. {\bf 02} (2000) 020 [hep-th/9912072].



\bibitem{inov} A. Iqbal, C. Vafa, N. Nekrasov and A. Okounkov,
{\it Quantum foam and topological strings}, J. High Energy Phys. {\bf 04} (2008) 011, [hep-th/0312022].




\bibitem{neova-kap} N. Nekrasov, H. Ooguri and C. Vafa, {\it S-duality and topological strings},
J. High Energy Phys. {\bf 10} (2004) 009, [hep-th/0403167]; A. Kapustin,
{\it Gauge theory, topological strings, and S-duality}, J. High Energy Phys. {\bf 09} (2004) 034,
[hep-th/0404041].



\bibitem{hsy-jhep09} H. S. Yang, {\it Emergent spacetime and the
origin of gravity}, J. High Energy Phys. {\bf 05} (2009) 012 [arXiv:0809.4728].



\bibitem{besse} A. L. Besse, {\it Einstein manifolds},  Springer-Verlag, Berlin, (1987).




\bibitem{griffiths-harris} P. Griffiths and J. Harris, {\it Principles of algebraic geometry},
John Wiley, New York (1978).




\bibitem{hsy-mirror} H. S. Yang, {\it Mirror symmetry in emergent gravity},
Nucl. Phys. B {\bf 922} (2017) 264 [arXiv:1412.1757].



\bibitem{pol-book} J. Polchinski, {\it String theory: Superstring theory and beyond},
Vol. I \& II, Cambridge Univ. Press, Cambridge (1998).




\bibitem{mnop} D. Maulik, N. Nekrasov, A. Okounkov and R. Pandharipande,
{\it Gromov-Witten theory and Donaldson-Thomas theory I},
Compositio Math. {\bf 142} (2006) 1263, [math.AG/0312059].



\bibitem{hsy-ijmp06} H. S. Yang, {\it Emergent gravity from noncommutative spacetime},
Int. J. Mod. Phys. A {\bf 24} (2009) 4473, [hep-th/0611174].



\bibitem{sw-darboux} L. Cornalba, {\it D-brane Physics and Noncommutative Yang-Mills Theory},
Adv. Theor. Math. Phys. {\bf 4} (2000) 271, [hep-th/9909081];
B. Jur\v co and P. Schupp, {\it Noncommutative Yang-Mills from equivalence of star products},
Eur. Phys. J. C {\bf 14} (2000) 367, [hep-th/0001032];
H. Liu, {\it $*$-Trek II: $*_n$ operations, open Wilson lines and the Seiberg-Witten map},
Nucl. Phys. B {\bf 614} (2001) 305, [hep-th/0011125].




\bibitem{symp-book} V. I. Arnold, {\it Mathematical methods of classical mechanics}, Springer (1978);
R. Abraham and J. E. Marsden, {\it Foundations of mechanics}, Addison-Wesley (1978).


	
\bibitem{beta-diff} T. Asakawa, H. Muraki and S. Watamura, {\it D-brane on Poisson manifold and Generalized Geometry},
Int. J. Mod. Phys. A {\bf 29} (2014) 1450089, [arXiv:1402.0942].



\bibitem{nc-sw} N. Seiberg and E. Witten, {\it String theory and noncommutative geometry},
J. High Energy Phys. {\bf 09} (1999) 032, [hep-th/9908142].



\bibitem{esw-hsy} H. S. Yang, {\it Exact Seiberg-Witten Map and Induced Gravity from Noncommutativity},
Mod. Phys. Lett. A {\bf 21} (2006) 2637, [hep-th/0402002]; R. Banerjee and H. S. Yang,
{\it Exact Seiberg-Witten map, induced gravity and topological invariants
in noncommutative field theories}, Nucl. Phys. B {\bf 708} (2005) 434, [hep-th/0404064].


\bibitem{hsy-eujc} H. S. Yang, {\it Noncommutative electromagnetism as a large N gauge theory},
Eur. Phys. J. C {\bf 64} (2009) 445, [arXiv:0704.0929].



\bibitem{hsy-cy} H. S. Yang, {\it Calabi-Yau manifolds from noncommutative Hermitian U(1) instantons},
Phys. Rev. D {\bf 91} (2015) 104002, [arXiv:1411.6115].


\bibitem{num-k3} M. Headrick and T. Wiseman, {\it Numerical Ricci-flat metrics on K3},
Class. Quantum Grav. {\bf 22} (2005) 4931, [hep-th/0506129].



\bibitem{hea} H. S. Yang, {\it Highly effective action from large N gauge fields},
Phys. Rev. D {\bf 90} (2014) 086006, [arXiv:1402.5134].



\bibitem{gi-u1} M. Salizzoni, A. Torrielli and H. S. Yang, {\it ALE spaces from noncommutative U(1)
instantons via exact Seiberg-Witten map}, Phys. Lett. B {\bf 634} (2006) 427, [hep-th/0510249];
H. S. Yang and M. Salizzoni, {\it Gravitational instantons from gauge theory},
Phys. Rev. Lett. {\bf 96} (2006) 201602, [hep-th/0512215];
H. S. Yang, {\it Instantons and emergent geometry}, Europhys. Lett. {\bf 88} (2009) 31002, [hep-th/0608013].



\bibitem{sly-prd} S. Lee, R. Roychowdhury and H. S. Yang, {\it Test of emergent gravity},
Phys. Rev. D {\bf 88} (2013) 086007, [arXiv:1211.0207];
{\it Topology change of spacetime and resolution of spacetime singularity in emergent gravity},
Phys. Rev. D {\bf 87} (2013) 126002, [arXiv:1212.3000].



\bibitem{nc-inst-ni} N. Nekrasov and A. Schwarz, {\it Instantons on noncommutative
$\mathbb{R}^4$, and (2, 0) superconformal six dimensional theory},
Commun. Math. Phys. {\bf 198} (1998) 689, [hep-th/9802068]; K. Furuuchi,
{\it Instantons on Noncommutative $\mathbb{R}^4$ and Projection Operators},
Prog. Theor. Phys. {\bf 103} (2000) 1043, [hep-th/9912047];
K. Y. Kim, B.-H. Lee and H. S. Yang, {\it Comments on instantons on noncommutative
$\mathbb{R}^4$}, J. Korean Phys. Soc. {\bf 41} (2002) 290, [hep-th/0003093].




\bibitem{sakoetal} A. Sako, {\it Instanton number of noncommutative U(n) gauge theory},
 J. High Energy Phys. {\bf 04} (2003) 023, [hep-th/0209139];
Y. Tian, C.-J. Zhu and X.-C. Song, {\it Topological charge of noncommutative ADHM instanton},
Mod. Phys. Lett. A {\bf 18} (2003) 1691, [hep-th/0211225].




\bibitem{hsy-jpcs12} H. S. Yang, {\it Towards a backround independent
quantum gravity}, J. Phys. Conf. Ser. {\bf 343} (2012) 012132, [arXiv:1111.0015].



\bibitem{nonins} M. Mari\~ no, R. Manasian, G. Moore and A. Strominger,
{\it Nonlinear instantons from supersymmetric p-branes},
J. High Energy Phys. {\bf 01} (2000) 005,  [hep-th/9911206].



\bibitem{lqg} C. Rovelli, {\it Loop quantum gravity}, Living Rev. Rel. {\bf 11} (2008) 5.



\bibitem{morita} B. Jur\v co, P. Schupp and J. Wess, {\it Noncommutative line bundle and Morita equivalence},
Lett. Math. Phys. {\bf 61} (2002) 171, [hep-th/0106110];
H. Bursztyn and S. Waldmann, {\it The characteristic classes of Morita equivalent star products
on symplectic manifolds}, Commun. Math. Phys. {\bf 228} (2002) 103, [math.QA/0106178].



\bibitem{ikkt} N. Ishibashi, H. Kawai, Y. Kitazawa and A. Tsuchiya, {\it A large-N reduced model as superstring},
Nucl. Phys. B {\bf 498} (1997) 467, [hep-th/9612115].



\bibitem{japan} H. Aoki, N. Ishibashi, S. Iso, H. Kawai, Y. Kitazawa and T. Tada,
{\it Noncommutative Yang-Mills in IIB matrix model}, Nucl. Phys. B {\bf 565} (2000) 176, [hep-th/9908141].




\bibitem{lebrun} C. LeBrun, {\it Explicit self-dual metrics on $\mathbb{C}\mathbb{P}_2 \sharp \cdots \sharp
\mathbb{C}\mathbb{P}_2$}, J. Diff. Geom. {\bf 34} (1991) 223.



\bibitem{bubble-geom} N. Bobev, B.E. Niehoff and N.P. Warner,
{\it Hair in the back of a throat: non-supersymmetric multi-center solutions from K\"ahler manifolds},
J. High Energy Phys. \textbf{10} (2011) 149, [arXiv:1103.0520];
N. Bobev, B.E. Niehoff and N.P. Warner, {\it New supersymmetric bubbles on $AdS_3 \times S^3$},
J. High Energy Phys. \textbf{10} (2012) 013, [arXiv:1204.1972].



\bibitem{bottom-up} S. Lee, R. Roychowdhury and H. S. Yang, {\it Notes on emergent gravity},
J. High Energy Phys. {\bf 09} (2012) 030, [arXiv:1206.0678].



\bibitem{riem-ym} J. J. Oh and H. S. Yang, {\it Einstein Manifolds As Yang-Mills Instantons},
 Mod. Phys. Lett. A {\bf 28} (2013), 1350097 [arXiv:1101.5185];
J. Lee, J. J. Oh and H. S. Yang, {\it An efficient representation of euclidean gravity I},
J. High Energy Phys. {\bf 12} (2011) 025, [arXiv:1109.6644].



\bibitem{ahep-yy} H. S. Yang and S. Yun, {\it Calabi-Yau Manifolds, Hermitian Yang-Mills instantons
and mirror yymmetry}, Adv. High Energy Phys. {\bf 2017} (2017) 7962426, [arXiv:1107.2095].



\bibitem{lcs} I. Vaisman, {\it Locally conformal symplectic manifolds}, Internat.
J. Math. \& Math. Sci. {\bf 8} (1985) 521;
D. Chinea, M. de Le��on and J. C. Marrero, {\it Locally conformal cosymplectic manifolds and timedependent
Hamiltonian systems}, Comment. Math. Univ. Carolin. {\bf 32} (1991) 383.



\bibitem{vaisman} I. Vaisman, {\it Lectures on the geometry of Poisson manifolds}, Birkh\"auser, Basel, (1994).



\bibitem{bfss} T. Banks,W. Fischler, S. H. Shenker and L. Susskind, {\it M theory as a matrix model: A conjecture},
Phys. Rev. D {\bf 55} (1997) 5112, [hep-th/9610043].



\bibitem{fuzzys2} J. Madore, {\it The Fuzzy Sphere}, Class. Quant. Grav. {\bf 9} (1992) 69;
H. Grosse, C. Klim\v c\' ik and P. Pre\v snajder, {\it Towards finite quantum field theory in noncommutative geometry},
Int. J. Theor. Phys. {\bf 35} (1996) 231, [hep-th/950]; U. Carow-Watamura and S. Watamura,
{\it Noncommutative geometry and gauge theory on fuzzy sphere},
Commun. Math. Phys. {\bf 212} (2000) 395, [hep-th/98011].




\bibitem{ourfuzzy} C.-T. Chan, C.-M. Chen, F.-L. Lin and H. S. Yang, {\it $CP^n$ model on fuzzy sphere},
Nucl.Phys. B {\bf 625} (2002) 327, [hep-th/0105087]; H. S. Yang and M. Sivakumar,
{\it Emergent geometry from quantized spacetime}, Phys. Rev. D {\bf 82} (2010) 045004, [arXiv:0908.2809].



\bibitem{georgi} H. Georgi, {\it Lie Algebras in Particle Physics: From Isospin to Unified Theories}, Advanced
Book Program, Addison-Wesley, Reading U.K. (1999).


\end{thebibliography}
\end{document}